\documentclass[prl,nobalancelastpage,twocolumn,nofootinbib,superscriptaddress,showpacs]{revtex4-1}

\usepackage{color,amsthm,amsmath,amsfonts,graphicx,bm}

\usepackage{setspace}
\usepackage{bm}
\usepackage{xcolor}
\usepackage{amsfonts}
\usepackage{amssymb}
\usepackage{amsmath}
\usepackage{epsfig}
\usepackage{graphicx}
\usepackage{enumerate}
\usepackage{multirow}
\usepackage{verbatim}
\usepackage{subfigure}
\usepackage{bbold}
\usepackage{soul}
\usepackage{scrextend}
\usepackage{bbm}

\usepackage{braket}
\usepackage{float}
\usepackage{hyperref}

\newcommand{\Tr}{\operatorname{Tr}}
\newcommand{\tr}{\operatorname{tr}}

\newcommand{\trf}{\operatorname{Tr}}
\newcommand{\trb}{\operatorname{tr}}

\newcommand{\mr}[1]{\mathrm{#1}}
\newcommand{\dg}{\dagger}

\newcommand{\fA}{f_1}
\newcommand{\fB}{f_2}
\newcommand{\fC}{f_3}
\newcommand{\fD}{f_4}
\newcommand{\nfill}{\nu}

\newcommand{\Deltac}{\Delta_{\mathrm{c}}}

\newcommand{\HoneD}{H_\mathrm{1D}}

\newcommand{\Hocc}{H_{\bar{\mathrm{c}}}}
\newcommand{\Hcc}{H_\mathrm{c}}

\newcommand{\tU}{\tilde{U}}

\newcommand{\otimesg}{ \otimes_{\mathrm{g}} }

\newcounter{notes}
\stepcounter{notes}

\graphicspath{{Figures/}}

\begin{document}

\title{Quantum Circuits Reproduce Experimental Two-dimensional Many-body Localization Transition Point}

\newcommand{\titleinfo}{Fermionic Quantum Circuits Reproduce Experimental Two-dimensional Many-body Localization Transition Point}

\date{\today}

\author{Joey Li}
\affiliation{Department of Physics and Institute for Condensed Matter Theory, University of Illinois at Urbana-Champaign, Urbana, Illinois 61801, USA}
\affiliation{Institute for Theoretical Physics, University of Innsbruck, 6020 Innsbruck, Austria}
\affiliation{Institute for Quantum Optics and Quantum Information of the Austrian Academy of Sciences, 6020 Innsbruck, Austria}
\author{Amos Chan}
\affiliation{Princeton Center for Theoretical Science, Princeton University, Princeton New Jersey 08544, USA}
\affiliation{Department of Physics, Lancaster University, Lancaster LA1 4YB, United Kingdom}

\author{Thorsten B. Wahl}
\affiliation{DAMTP, University of Cambridge, Wilberforce Road, Cambridge, CB3 0WA, United Kingdom}

\begin{abstract}
While many studies point towards the existence of many-body localization (MBL) in one dimension, the fate of higher-dimensional strongly disordered systems is a topic of current debate. The latest experiments as well as several recent numerical studies indicate that such systems behave many-body localized -- at least on practically relevant time scales. However, thus far, theoretical approaches have been unable to quantitatively reproduce experimentally measured MBL features -- an important requirement to demonstrate their validity. In this work, we use fermionic quantum circuits as a variational method to approximate the full set of eigenstates of two-dimensional MBL systems realized in fermionic optical lattice experiments. Using entanglement-based features, we obtain a phase transition point in excellent agreement with the experimentally measured value. Moreover, we calculate, the filling fraction-dependent MBL phase diagram, an important feature which has not been addressed in previous literature. We argue that our approach best captures the underlying charge-density-wave experiments and compute the mean localization lengths, which can be compared to future experiments. 
\end{abstract}

\maketitle

\textit{Introduction. ---} 
An important requirement for the laws of thermodynamics to be meaningful is that closed systems act as their own heat baths. For quantum many-body systems, this requirement is expressed in the eigenstate thermalization hypothesis~\cite{deutsch1991quantum,srednicki1994chaos} (ETH). However, with the discovery of many-body localization~\cite{Fleishman1980,gornyi2005interacting,basko2006metal} (MBL), it became apparent that not only fine-tuned systems are able to evade the ETH~\cite{anderson1958absence}, but that there exists an entire phase of matter that lacks thermalization. For one-dimensional strongly disordered quantum systems, the existence of MBL is underpinned by various numerical studies~\cite{znidaric2008many,pal2010mb,Bardarson2012,kjall2014many}, experimental observations~\cite{Schreiber842,Lukin2018,Smith_MBL,Roushan2017,Choi2017}, and a recent rigorous proof using only very mild assumptions~\cite{imbrie2016many}.

\begin{figure}[h]
\centering
\includegraphics[width=0.48 \textwidth  ]{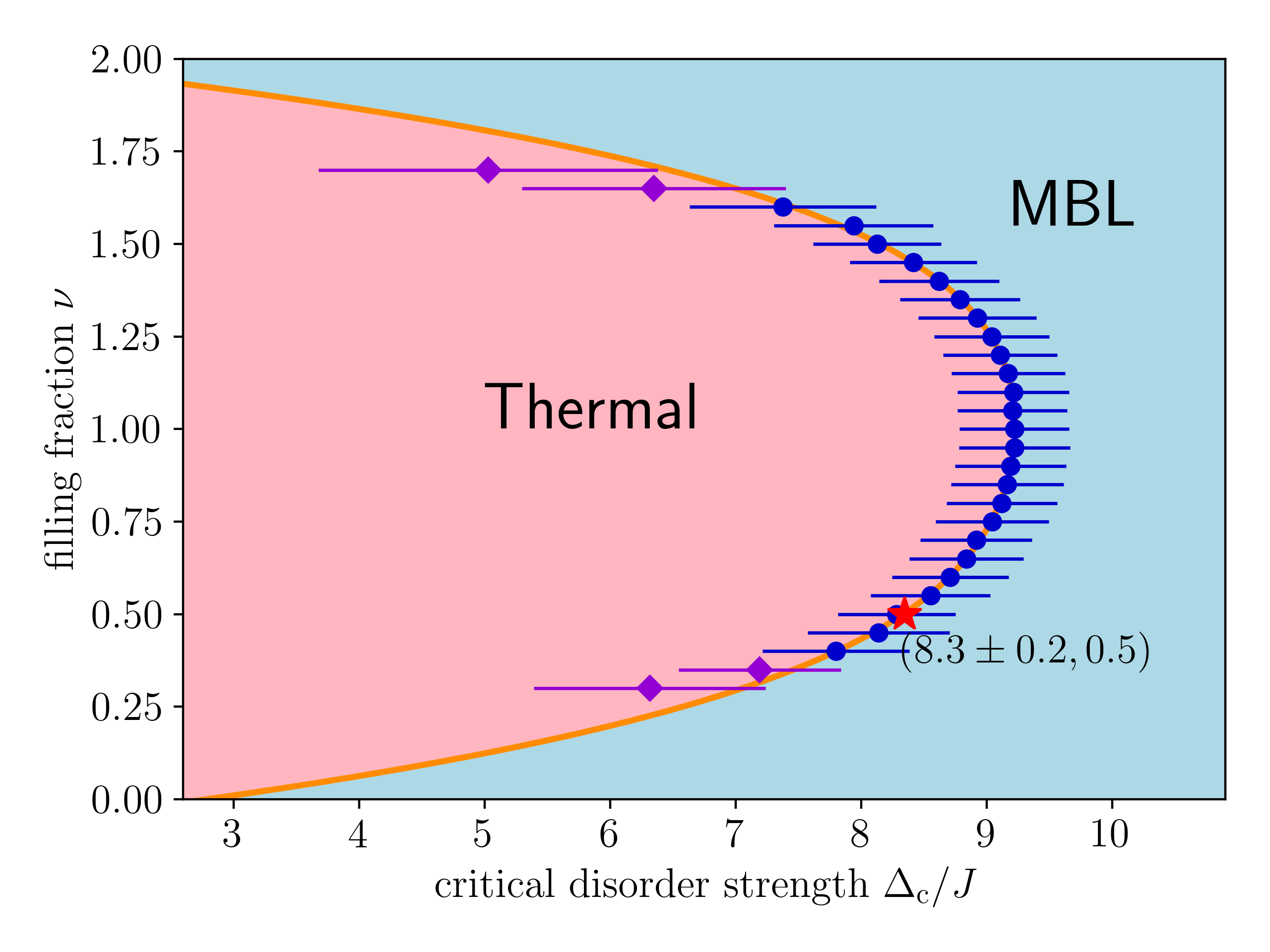} 
    \caption{Filling fraction-dependent phase diagram. We plot $\Delta_c$, with error bars indicating $\pm 1$ standard error, as a function of filling fraction from $\nu=0.3$ to $\nu=1.7$. The solid line is a fourth order polynomial fit using only the $\nu\in[0.4,1.6]$ data points (blue dots). The red star corresponds to the expected phase transition point at the experimental filling fraction of $\nu = 0.5$.}
     \label{fig:phasediagram}
\end{figure}

In two dimensions, MBL would allow for topologically-protected quantum computation away from low temperatures~\cite{2013Bauer_Nayak,Huse2013LPQO,bahri2015localization,Wahl2020,Venn2023}, a promising feature as we enter the era of noisy intermediate-scale quantum computing~\cite{Arute2019googlesycamore}. However, whether MBL exists beyond one dimension is still an important open question:  Experimental observations are supportive of higher-dimensional MBL~\cite{Choi1547,bordia2017quasiperiodic2D,2D_quantum_bath}, and several theoretical studies are also in favor of this scenario~\cite{Thomson2018,2DMBL,Kennes2018,Alet2D,DeTomasi2019,Kshetrimayum2020,Chertkov2021,Tang2021,Decker2021,Burau2021,Kim2021}, while others plead against it~\cite{deRoeck2017Stability,Gopalakrishnan2019,Doggen2020}. A possible explanation of these divergent results is that strongly disordered quantum systems in higher dimensions eventually thermalize, but only after extremely long times~\cite{chandran2016higherD,Gopalakrishnan2019}. As a result, it is likely that such systems behave many-body localized on all experimentally relevant time scales. Hence, the practically most relevant task is the description of the phenomenon of higher-dimensional MBL on those time scales.

In this work, we use shallow fermionic two-dimensional quantum circuits to simulate the MBL-to-thermal transition observed in optical lattice experiments with fermions subject to a quasi-periodic potential~\cite{bordia2017quasiperiodic2D}. We obtain a phase transition point in excellent agreement with the one measured experimentally and argue that this is due to the corresponding charge-density-wave initializations being well captured by shallow quantum circuit simulations. We furthermore calculate the filling fraction-dependent MBL phase diagram (shown in Fig.~\ref{fig:phasediagram}), an important feature which has been neglected in previous experimental and theoretical studies. Throughout, we use the semantics of a phase transition (although strictly speaking it might be a crossover~\cite{Gopalakrishnan2019}), as we focus on the behavior on experimentally relevant time scales, which is exactly what our method is able to capture. 

\begin{figure}[t]
\centering
\begin{picture}(250,455)
\put(0,0){\includegraphics[width=0.48 \textwidth  ]{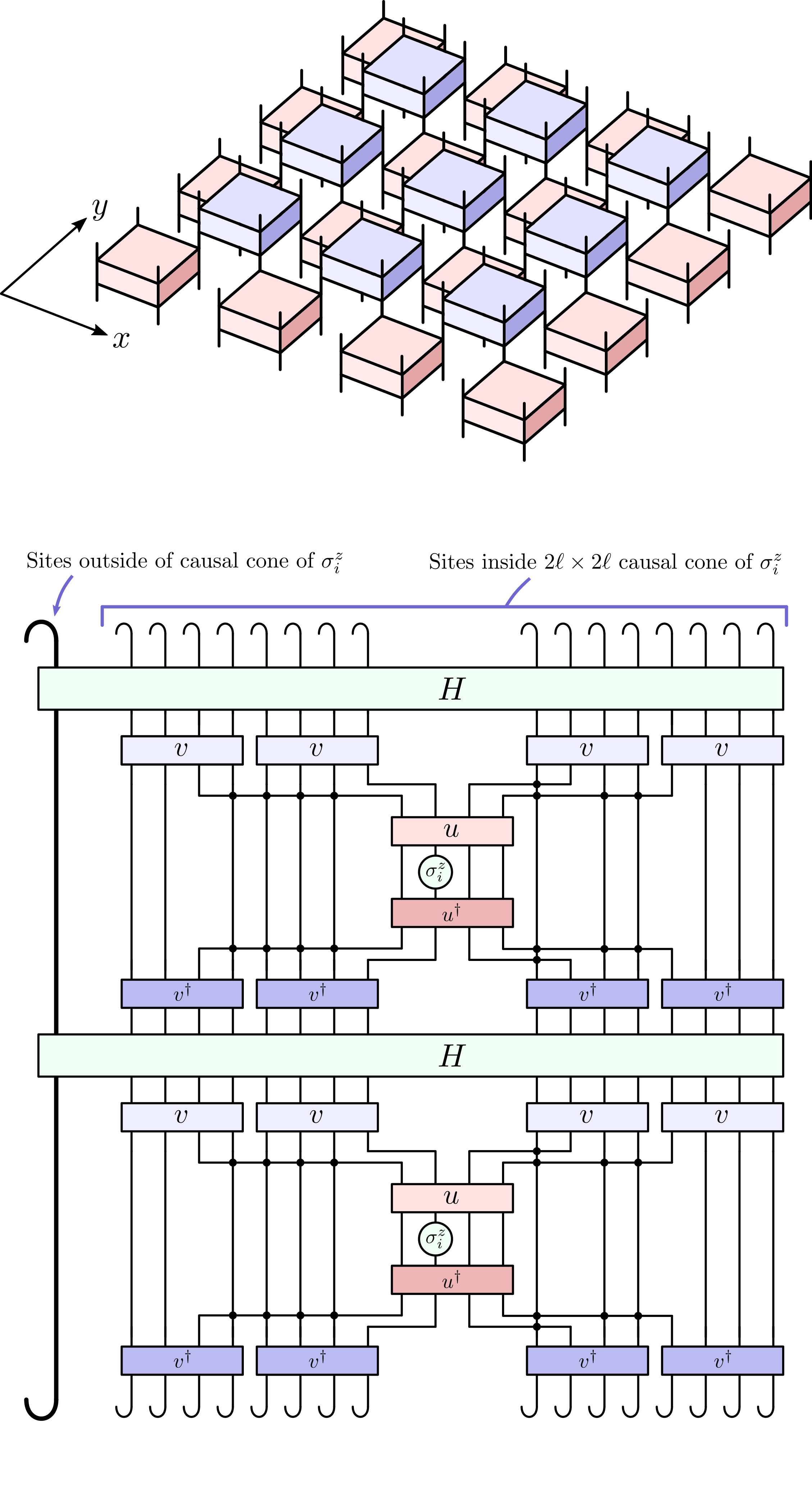}}
\put(0,447){\textbf{(a)}}
\put(0,292){\textbf{(b)}}
\end{picture}
    \caption{
    (a) The approximate eigenstates are contained in the variational ansatz $\tilde U$, a tensor network composed of two layers of unitary gates arranged in a brick-wall geometry. Each line in this tensor network has dimension $d=4$, the on-site Hilbert space dimension of the model Eq.~\eqref{eq:hamiltonian}.
    (b) $\tilde U$ is optimized by minimizing $-\Tr(H\tilde \tau^z_{i,\mu} H \tilde \tau^z_{i,\mu})$ (illustrated, as a tensor network contraction), which depends only on a $4 \times 4$ causal cone around $\sigma_{i,\mu}^z$. See~\cite{supplementary} for details on how this diagram is derived and calculated. 
    } \label{fig:circuit}
\end{figure}

A quantum circuit is a representation of a unitary acting on many sites in terms of smaller unitaries (also called gates) acting non-trivially only on small patches of sites each. If the gates can be applied in parallel, it is possible to stack them in a brickwork structure, as in Fig.~\ref{fig:circuit}a. If only a small number of layers of gates is required for such a brickwork structure, we call the quantum circuit shallow. Shallow quantum circuits have proven to be a powerful tool, both for the numerical simulation of MBL~\cite{Pollmann2016TNS,Wahl2017PRX,2DMBL,Wahl2022,Venn2023} and the analytical classification of symmetry-protected topological MBL phases~\cite{Thorsten,1DSPTMBL,2DSPTMBL}. As opposed to other approaches, quantum circuits approximate the full set of eigenstates in the MBL regime and give access to all their local observables. In one dimension, this feature allows one to reproduce the MBL-to-thermal transition point~\cite{Wahl2017PRX} found in exact diagonalization studies~\cite{Luitz2015}, even with very simple quantum circuits. While previous two-dimensional quantum circuit simulations displayed striking features of an MBL phase in the strongly disordered regime~\cite{2DMBL}, they significantly overestimated the phase transition point found in the corresponding experiments~\cite{Choi1547}. 
These were initialized with one half of the optical trap filled, whereas the fermionic optical lattice experiments of Ref.~\cite{bordia2017quasiperiodic2D} were initialized with charge-density-waves. This gives quantum circuit simulations of those fermionic experiments two important advantages over the simulation of bosonic experiments: (i) The numerical simulations do not require a truncation of the on-site Hilbert space dimension. (ii) Charge-density-waves require only few hops of the atoms to reach thermal equilibrium, a process more amenable to a description by shallow quantum circuits. 
We note that fermionic quantum circuit simulations do not suffer from the sign problem plaguing Monte Carlo approaches. 
We calculate the transition point as a function of the filling fraction, reproducing the experimental result of Ref.~\cite{bordia2017quasiperiodic2D} at half filling. We also compute the average localization length as a function of disorder strength, which can be measured in future domain-wall experiments~\cite{Choi1547}.

\textit{Formalism. ---}  In one dimension, MBL systems have been shown to possess a complete set of local integrals of motion~\cite{serbyn2013local,Huse_MBL_phenom_14,chandran2015constructing,ros2015integrals,Inglis_PRL2016,Rademaker2016LIOM,Monthus2016,Goihl2018,Abi2017} (LIOMs), also known as l-bits. For a spin-$1/2$ chain, they are typically denoted as $\tau_i^z$ and by definition commute with the MBL Hamiltonian $\HoneD$ and with each other, $[\HoneD, \tau_i^z] = [\tau_i^z,\tau_j^z] = 0$. The LIOMs are related via a local unitary~\cite{Chen2013} $U$ (which also diagonalizes the Hamiltonian) to the on-site spin-$z$ Pauli operators, $\tau_i^z = U \sigma_i^z U^\dg$. As a result, $\tau_i^z$ is exponentially localized around site $i$, where the corresponding characteristic decay length is known as the localization length $\xi_L$. The Hamiltonian can be expressed in terms of LIOMs, $\HoneD = c + \sum_i c_i \tau_i^z + \sum_{i,j} c_{ij} \tau_i^z \tau_j^z + \sum_{i,j,k} c_{ijk} \tau_i^z \tau_j^z \tau_k^z + \ldots$, where $|c_{ijk\ldots}|$ decays exponentially with the maximum distance between the indexed sites. 

In two dimensions, any set of local quantities $\{\tau_i^z\}$ may not exactly commute with the Hamiltonian; however, the inverse of the commutator in operator norm $\|[H_{2D},\tau_i^z]\|_\mr{op}$ serves as a lower bound on relaxation times~\cite{chandran2016higherD}. Hence, systems with a very small commutator behave many-body localized on very long time scales~\cite{chandran2016higherD,Gopalakrishnan2019}, giving rise to the experimental observation of two-dimensional MBL. Setting aside the question whether MBL persists in the infinite-time limit, here we use fermionic quantum circuits to describe the MBL-to-thermal transition found in optical lattice experiments with fermions~\cite{bordia2017quasiperiodic2D}. To that end, we define a variational quantum circuit $\tilde U$ which approximately diagonalizes the Hamiltonian, and use it to construct a set of approximate LIOMs $\tilde \tau_i^z = \tilde U \sigma_i^z \tilde U^\dg$~\cite{2DMBL}. We optimize the unitaries constituting $\tilde U$ by minimizing the commutator with the Hamiltonian in fermionic trace norm $\sum_i \|[H,\tilde \tau_i^z]\|_\mr{tr}^2$. Since the approximate LIOMs we will obtain will commute very well with the Hamiltonian, they will correspond to local observables, which equilibrate only on very long time scales~\cite{chandran2016higherD}. This enables us to describe the MBL-like behavior seen in experiments. The likely transience of MBL in two dimensions implies that the transition seen in experiments is actually a (relatively abrupt~\cite{Gopalakrishnan2019}) crossover between a quickly thermalizing regime and a regime where thermalization takes place on non-accessible time scales. It is currently unclear whether the arguments in favor of metastable MBL also apply to quasi-periodic systems~\cite{deRoeck2017Stability,Singh2021}. 

\textit{Model and numerical approach. ---} We study the model Hamiltonian underlying the fermionic optical lattice experiments of Ref.~\cite{bordia2017quasiperiodic2D}.  The system consists of spin-$1/2$ fermions on a two-dimensional $N \times N$ square lattice governed by the Fermi-Hubbard Hamiltonian with a quasi-periodic potential, 
\begin{align}\label{eq:hamiltonian}
    H &= -J\sum_{\langle i,j \rangle} \sum_{\mu = \uparrow,\downarrow} (c^\dagger_{i,\mu}c_{j,\mu} + c_{j,\mu}^\dg c_{i,\mu}) +  V \sum_i  n_{i,\uparrow}n_{i,\downarrow} \notag \\
    &+ \Delta\sum_i \big[\cos(2\pi\beta_x x)+\cos(2\pi\beta_y y)\big](n_{i,\uparrow} + n_{i,\downarrow}),
\end{align}
where the lattice sites are labeled by $i = (x,y)$. $c^\dagger_{i,\mu},\, c_{i,\mu}$ are fermionic creation and annihilation operators with spin-$z$ component $\mu = \uparrow, \downarrow$, and $n_{i,\mu} = c^\dagger_{i,\mu}c_{i,\mu}$ is the particle number operator.  Using the experimental model parameters, we set $\beta_x = 0.721$ and $\beta_y = 0.693$, and the on-site interaction strength to $V = 5J$. $\Delta$ is the disorder strength; experimentally the MBL phase was observed above the critical disorder strength $\Delta_c^\mr{exp} = 9 \pm 0.5 J$.

In our computational approach, we consider multiple disorder realizations of a $10\times 10$ system with periodic boundary conditions to emulate the properties of the experimental system in the center of the optical trap.  To produce different disorder realizations, we shift the quasi-periodic potential by a random amount, i.e. we use the potential $\Delta[\cos(2\pi(\beta_x x + \delta x))+\cos(2\pi(\beta_y  y + \delta y ))]$  with $\delta x , \delta y \in [0,1)$ drawn from a uniform distribution, where we keep the same choices for different $\Delta$.
We introduce the variational quantum circuit $\tilde U$, which is composed of smaller unitary operators arranged in two layers, as depicted in Fig.~\ref{fig:circuit}a.  The small unitaries each act on $2\times 2$ patches of sites and are restricted to be real and to preserve particle number (as the overall Hamiltonian does), which also ensures they have even parity.

In order to optimize $\tilde U$, we define a cost function $f(\tilde U)$ such that $f(\tilde U)=0$ if $\tilde U$ diagonlizes the Hamiltonian.  $f(\tilde U)$ may be written as~\cite{supplementary,2DMBL}
\begin{align}
    f(\tilde U) 
    = \frac{1}{N^24^{N^2}}\sum_i \sum_{\mu = \uparrow,\downarrow} \big[ \trf(H^2) - \trf(H\tilde \tau_{i,\mu}^z H \tilde \tau_{i,\mu}^z)\big],~\label{eq:fom}
\end{align}
which consists of a constant term $\Tr(H^2)$ and a nontrivial term whose tensor network representation is shown in Fig.~\ref{fig:circuit}b. 
Here, we have to account for the fact that the on-site Hilbert space dimension of the model is 4, i.e., there are two approximate LIOMs per site $i$, $\tilde \tau_{i,\mu}^z(\tilde U) = \tilde U \sigma^z_{i,\mu} \tilde U^\dg$, where $\sigma^z_{i,\uparrow}$ acts on site $i$ as $\sigma^z \otimes \mathbb{1}$ and $\sigma^z_{i,\downarrow}$ as $\mathbb{1} \otimes \sigma^z$. Expression~\eqref{eq:fom} can be decomposed into a sum of local parts whose complexity is independent of the system size, allowing for the efficient calculation of the cost function and numerical optimization of the unitaries using automatic differentiation~\cite{supplementary}.  %

\begin{figure}
\centering
	\begin{picture}(250,330)
		\put(0,166){\includegraphics[width=0.46\textwidth  ]{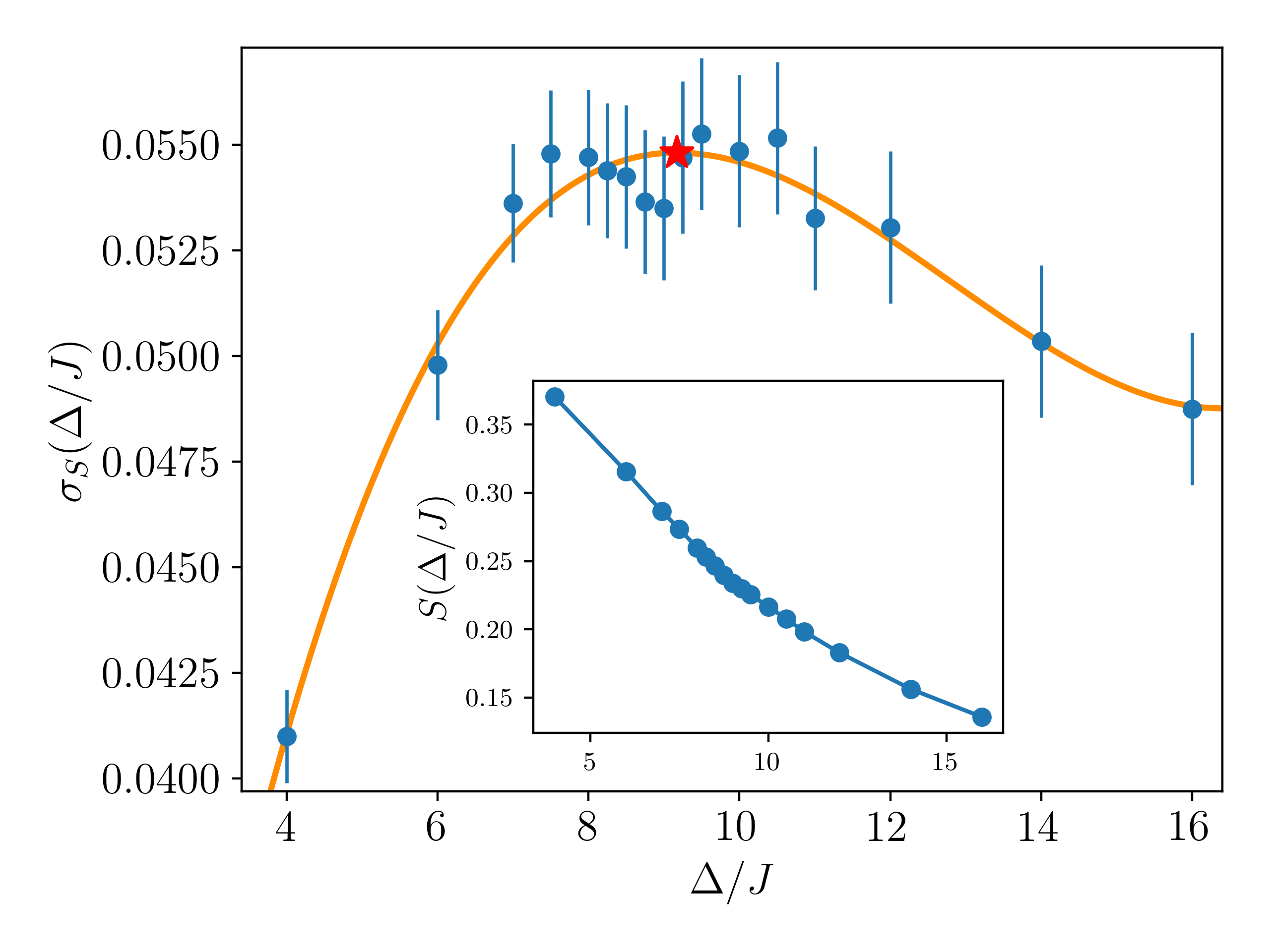}} 
        \put(0,0){\includegraphics[width=0.46 \textwidth  ]{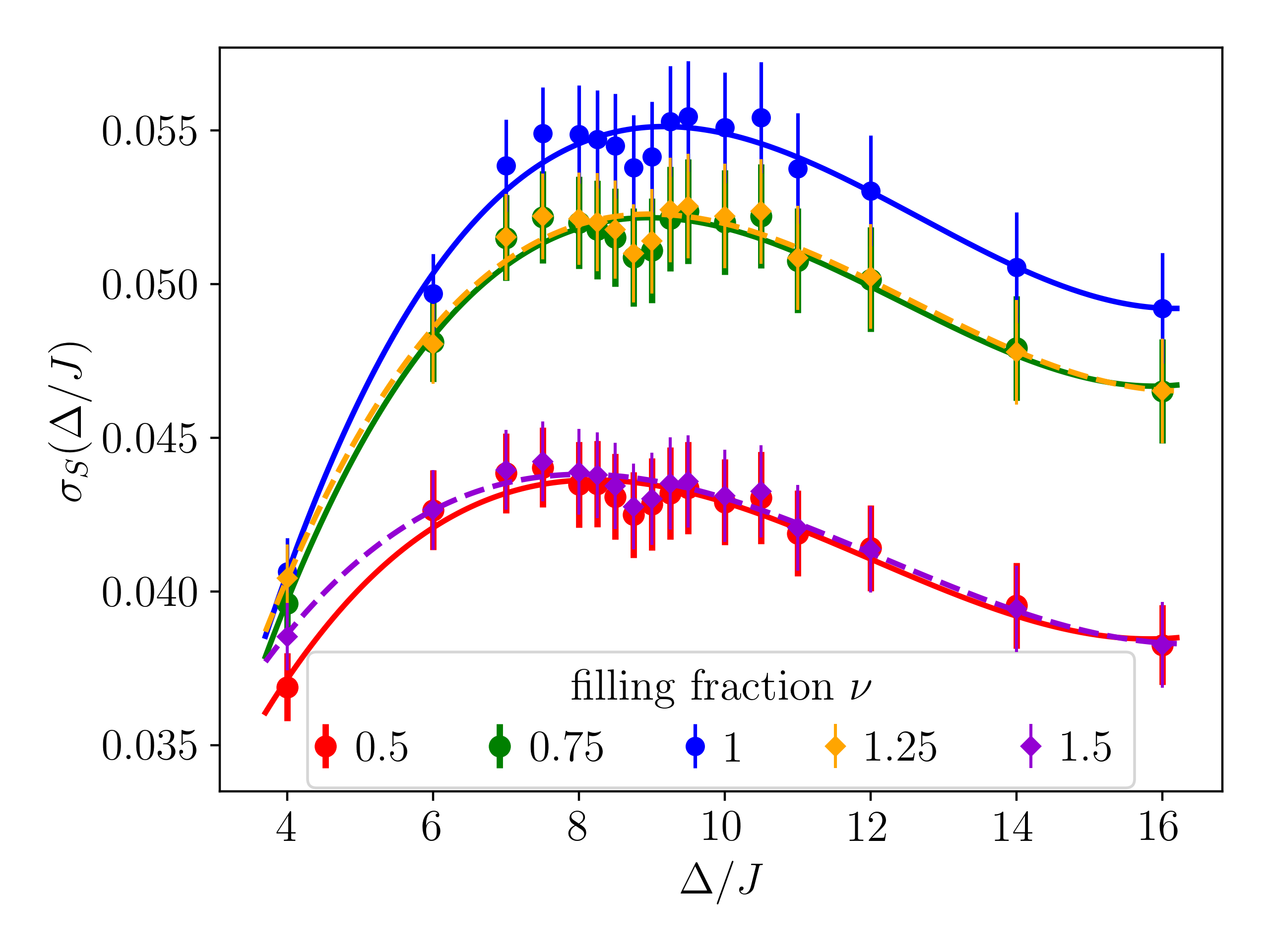}}
        \put(0,312){\textbf{(a)}}
        \put(0,152){\textbf{(b)}}
    \end{picture}
    \caption{(a) Standard deviation, over 10 disorder realizations, of the entanglement entropy (averaged over $1000$ approximate eigenstates) as a function of $\Delta/J$, averaged over sites. The curve is a third order polynomial fit and error bars denote two standard errors (for the error analysis, see~\cite{supplementary}). The peak of this curve, located at $(9.2,0.055)$ and marked by a red star, indicates the transition point of the overall Hamiltonian. Inset: Entanglement entropy $S$ (averaged over sites, eigenstates, and disorder realizations) as a function of $\Delta/J$.
    (b) Same plot as in (a) but with filling-fraction resolution, for a few representative values of $\nu$, each averaged over $10^4$ approximate eigenstates. 
    }
    \label{fig:entropyData}
\end{figure}

\textit{Results. ---}
We determined the location of the MBL-to-thermal transition of this model by computing the on-site entanglement entropy $S_i$ averaged over approximate eigenstates from the optimized $\tilde U$ as a function of disorder strength $\Delta$. Even though the Fock space dimension grows exponentially with the system size, this can be done in an efficient fashion using the approach outlined in \cite{supplementary, 2DMBL}. 

The standard deviation $\sigma_{S,i}$ of the averaged entanglement entropy with respect to disorder realizations peaks at the transition point~\cite{kjall2014many}. To reduce statistical fluctuations, we average $\sigma_{S,i}$ over site positions $i$~\cite{Wahl2017PRX,2DMBL}, resulting in the behavior shown in Fig.~\ref{fig:entropyData}a. We observe a maximum at $\Delta_c \approx 9.2 J$. 
This value corresponds to an average over 1000 randomly chosen approximate eigenstates per disorder realization and all sites $i$. 
In~\cite{supplementary}, we analyse the probability distributions of entanglement entropies, which display characteristic multimodal features near $\Delta = 9J$, forming another signature of the transition.

The filling fraction $\nfill$ of almost all eigenstates is near its mean value of 1. 
However, the charge-density-wave evolutions of Ref.~\cite{bordia2017quasiperiodic2D} only involved eigenstates with filling fraction $\nfill = 0.5$. Since lower filling fractions are associated with lower energies and the transition point depends on the energies of the involved eigenstates~\cite{Luitz2015,2DMBL,Tang2021}, our results need to be specified for filling fraction $\nfill = 0.5$ in order to allow for an appropriate comparison with the experimental result. To that end, we carried out the above analysis by resolving the maximum of the site-averaged standard deviation of the entanglement entropy for different fillings.  
We show the filling fraction-resolved transition point $\Delta_c(\nfill)$ in 
Fig.~\ref{fig:phasediagram}. The transition points were obtained from the maxima of the filling fraction-resolved entanglement entropy fluctuations of Fig.~\ref{fig:entropyData}b. The error bars were obtained by randomly varying the data points of Fig.~\ref{fig:entropyData}b according to normal distributions of variances equal the corresponding standard errors~\cite{supplementary}. 

The result at half filling based on a statistical error analysis~\cite{supplementary} is $\Delta_c(0.5) = 8.3 \pm 0.4 J$ ($\pm 2$ standard deviations, i.e., $95 \%$ confidence interval). However, there is also a systematic error due to the finite entanglement inherent to our tensor network ansatz. We estimate this error by comparing to a truncated model of Eq.~\eqref{eq:hamiltonian} that excludes double occupancies, which are relatively rare at filling fraction $\nu = 0.5$. 
Crucially, this restriction to on-site Hilbert space dimension $d = 3$ implies that all the constituting unitaries have local dimension $d = 3$ (bond dimension); thus, studying this truncated model effectively corresponds to studying the original model with a tensor network ansatz of lower bond dimension. 
We performed the above-described process (optimizing unitaries and then observing entanglement fluctuation) for the truncated model, obtaining a phase transition point of $\Delta_c^{d = 3}(0.5) = 5.3J$~\cite{supplementary}. However, this value should be scaled before comparing with the result from the un-truncated model. In the truncated model at half filling, each fermion is on average surrounded by two fermions, one spin-up and one spin-down. However, the exclusion of double occupancies prevents the central fermion from hopping to the neighboring site occupied with opposite spin, effectively reducing the hopping strength (very roughly) by a factor of $3/4$, i.e. $J\rightarrow J_\text{eff} = (3/4)J$, so that $\Delta/J\rightarrow\Delta/J_\text{eff}=(4/3)(\Delta/J)$. Hence, a better estimate of the transition point predicted by bond dimension $3$ calculations (as opposed to actual truncation of the physical model) would be ${\Delta_c'}^{d=3}(0.5) \approx (4/3)\Delta_c^{d=3}(0.5) \approx 7 J$, close to $\Delta^{d=4}_c(0.5) = 8.3 \pm 0.4 J$. We also note that in the limit $\nu \rightarrow 0$, where double occupancies can be safely neglected, the original phase transition points for $d = 3,4$ are close. We obtained $\Delta_c^{d=3}(\nu \rightarrow 0) = 4.3J$~\cite{supplementary} and $\Delta_c^{d=4}(\nu \rightarrow 0) \approx 3J$ (see the intersection of the critical line with the horizontal axis in Fig.~\ref{fig:phasediagram}), moving towards the exact, non-interacting result of $\Delta_c^\mr{non-int}(\nu \rightarrow 0) = 2J$. For more details, see~\cite{supplementary}. Similarly, the phase transition point at $\nu = 0.5$ increases with increasing bond dimension, i.e., it drifts closer to the experimental value. We cannot specify a precise error bar associated with this drift, though the corresponding error combined with the above statistical error is very likely to overlap with the range found experimentally, $\Delta_c^\mr{exp} = 9 \pm 0.5 J$~\cite{bordia2017quasiperiodic2D}.

We now address the question as to why no such agreement was found in the earlier study of Ref.~\cite{2DMBL}, attempting to describe the ``half-moon'' experiments of Ref.~\cite{Choi2017}. We first note that our method describes a set of approximate LIOMs $\tilde \tau_i^z$ (rather than the actual integrals of motion of the Hamiltonian $H$), which due to the short depth of the quantum circuit act non-trivially only on a finite patch of sites.  Our method thus captures the phase transition point of a fiducial Hamiltonian $\tilde H$ built only of sums of products of the $\tilde \tau_i^z$, such that $\tilde H$ is exactly diagonalized by our quantum circuit $\tilde U$, which only approximately diagonalizes $H$ itself. Therefore, we can choose $\tilde H \approx H$. This relation can be used to show~\cite{supplementary} that higher-order products of the $\tilde \tau_i^z$ in $\tilde H$ vanish. This in turn implies that in a time evolution $\tilde H$ can propagate a domain wall only across a fixed number of sites $\mathcal{O}(\ell)$, where $\ell=2$ is the size of the small unitaries. Therefore, our quantum circuit describes a Hamiltonian which is able to equilibrate a charge-density-wave, but not a ``half-moon'' configuration.  Our approach therefore captures the earlier type of experiments, but not the latter.

\begin{figure}[t]
\centering
\includegraphics[width=0.45 \textwidth  ]{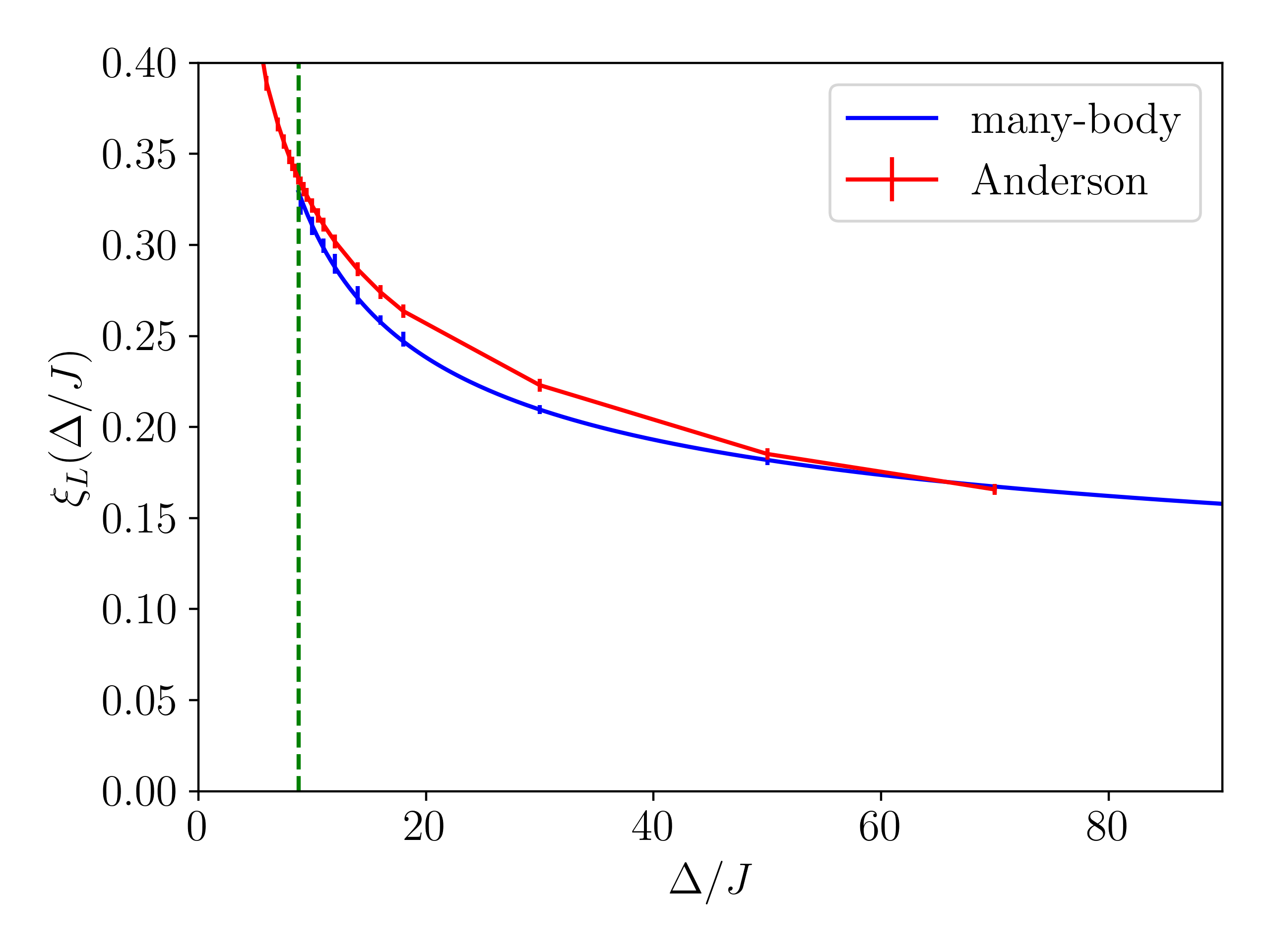}
    \caption{Localization length $\xi_L$ (averaged over lattice sites and disorder realizations) as a function of $\Delta/J$ with a $1/\ln(\Delta/J)$ fit (blue)~\cite{Hundertmark,Pietracaprina2016,Scardicchio2017} and the non-interacting result $V=0$ (red) shown. The error bars correspond to fluctuations between disorder realizations. The green dashed line indicates the transition point $\Delta_c = 9.2J$ found for the overall Hamiltonian. $\xi_L$ does not diverge at the transition point, a universal problem of numerical approaches~\cite{Abi2017}. The localization length coincides with the non-interacting result within uncertainties for large $\Delta$, but starts to deviate already for $\Delta < 50J$. 
    } \label{fig:localizationlength}
\end{figure}

We also calculated the average localization length $\xi_L$ as a function of disorder strength as a quantity which can be compared with future domain-wall experiments with fermions~\cite{Choi1547}. To that end, we adapted the weight function $w_i(\mathcal{O})$ introduced in Ref.~\cite{Abi2017} measuring the weight of a Hermitian operator $\mathcal O$ on site $i$ to fermionic spin-$1/2$ systems,
\begin{align}
    w_i(\mathcal{O}) = {\sum_{\beta,\gamma}}' \Tr\left[\left(\mathcal{O} - \hat\sigma_{i,\uparrow}^\beta \hat \sigma^\gamma_{i,\downarrow}\mathcal{O} \hat \sigma^\gamma_{i,\downarrow} \hat\sigma^\beta_{i,\uparrow}\right)^2\right], \label{eq:weight}
\end{align}
where $\hat \sigma_{i,\mu}^x \equiv c_{i,\mu} + c_{i,\mu}^\dg$, $\hat \sigma_{i,\mu}^y \equiv -i c_{i,\mu} + i c_{i,\mu}^\dg$, $\hat \sigma_{i,\mu}^z \equiv 1 - 2c_{i,\mu}^\dg c_{i,\mu}$, and $\hat\sigma_{i,\mu}^0 = \mathbb{1}$. 
The $\sum'$ sum is defined over $\beta, \gamma \in \{0,x,y,z\}$, but omitting the term $\beta = \gamma = 0$. $w_i(\tilde \tau_{j,\mu}^z)$ can be calculated efficiently~\cite{supplementary}. Within the causal cone of $\tilde \tau_{j,\mu}^z$ it decays exponentially with the distance between sites $i$ and $j$. The corresponding decay length averaged over all sites $i,j$ and disorder realisations yields $\xi_L$ for a given disorder strength $\Delta$. This is plotted in Fig.~\ref{fig:localizationlength} and compares well to the Anderson localization lengths (shown as well) in the limit of large $\Delta$. 

\textit{Conclusions. ---} We used a fermionic quantum circuit approach to simulate the two-dimensional MBL phase observed in the optical lattice experiments of Ref.~\cite{bordia2017quasiperiodic2D}. A careful analysis of the entanglement features for different filling fractions yields an MBL-to-thermal transition point in excellent agreement with the experimental result.  Our obtained phase transition point as a function of filling fraction can be compared to future charge-density-wave experiments with occupied columns away from one fermion per site. We also calculated the average localization length as a function of disorder strength, another quantity which can be compared to future experiments. Finally, we provided an argument why quantum circuits built of unitaries acting on patches of $2 \times 2$ sites best capture the phase transition point found with charge-density-wave initializations, explaining the discrepancy between the phase transition points found in Refs.~\cite{Choi1547} and~\cite{2DMBL}. A promising direction for future research is to increase the size $\ell$ of the small unitaries composing $\tilde U$ ($\ell=2$ in the present work) and analyze if the resulting phase transition points come closer to the result of half-moon experiments. Larger unit cells might also make it possible to test if regions of anomalously small disorder delocalize their environments, leading to the eventual breakdown of MBL in two dimensions~\cite{deRoeck2017Stability, Altman2018stability}.

\textit{Acknowledgments. ---} We are grateful to David Huse, Rahul Nandkishore, and Steven Simon for useful discussions. 
JL is supported by ERC Starting grant QARA (Grant No.~101041435). AC is supported by fellowships from the Croucher Foundation and the PCTS at Princeton University. TBW acknowledges support through the ERC Starting Grant No.~678795 TopInSy.

\vspace{11pt}

\bibliography{biblio}{}

\newpage
\onecolumngrid

\appendix

\begin{center}
\Large{Supplemental Material\\Quantum Circuits Reproduce Experimental Two-dimensional Many-body Localization Transition Point}
\end{center}
\setcounter{equation}{0}
\setcounter{figure}{0}
\renewcommand{\thetable}{S\arabic{table}}
\renewcommand{\theequation}{S\arabic{equation}}
\renewcommand{\thefigure}{S\arabic{figure}}
\setcounter{secnumdepth}{0}

\vspace{-20pt}
\section{1. Diagrammatic approach to fermionic tensor networks}\label{sec:fermion_diag}

Here we review the idea of super vector spaces following Refs.~\cite{susy2004, 1DSPTMBL, vers1610}, and the diagrammatic approach to fermionic tensor networks introduced in Ref.~\cite{1DSPTMBL} (see related but different approaches in Refs.~\cite{guif200910, guif201001, gu201004, ign201005, orus2010}). 
A super vector space $V = V^0 \oplus V^1$ is a direct sum of the vector spaces $V^0$ and $V^{1}$ containing even and odd parity vectors, respectively. 
A vector  $\ket{a} \in V$ is homogeneous if it belongs either to $V^0$ or $V^1$, and its parity is denoted by $|a| \in \{0,1\}$, 0 for even and 1 for odd. $V^*$ denotes the dual super vector space. 
The graded tensor product of two homogeneous vectors $\ket{a_1}$ and $\ket{a_2}$ is $\ket{a_1} \otimesg \ket{a_2} \in V \otimesg  V$, and its parity is  $|a_1| + |a_2| \mod 2$. 
The reordering of vectors 
within a graded tensor product is the isomorphism
$
\mathcal{F}:  \ket{a_1}  \otimesg \ket{a_2} \to (-1)^{|a_1| |a_2|} \ket{a_2}  \ \otimesg \ket{a_1 } $. 
The reordering of graded tensor products in $V^* \otimesg W$, $V \otimesg W^*$ and $V^* \otimesg W^*$ is similarly defined. 
The contraction $\mathcal{C}$ is the homomorphism 
$
\mathcal{C}:
\bra{a_1  }  \otimesg \ket{a_2 } \to \braket{a_1 | a_2}$. 
An operator acting on the super vector space $V$ is 
$
\hat{O} = \sum_{a,b} O_{a,b} \ket{a} \otimesg \bra{b} \quad \in V \otimesg V^* , 
$
which has parity $|\mathrm{O}| := |a| + |b| \mod 2$ (where $|a\rangle$ and $|b\rangle$ denote basis vectors with a fixed parity). Higher rank operators are similarly defined.

	\begin{figure}[b]
	\centering
	\includegraphics[width=0.8\columnwidth]{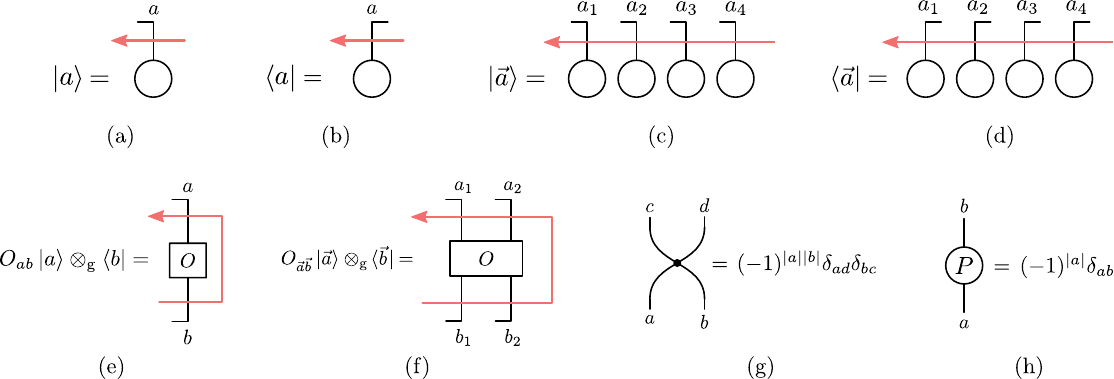}
	\caption{Diagrammatic representation of 
	(a) a ket $\ket{a} $;
	(b) a bra $\bra{a} $;
	(c) a many-body ket $\ket{\vec{a}}$;
	(d) a bra $\bra{\vec{a}}$ in the standard ordering;
	(e) a rank-2 operator $\hat{O}$; 
	(f)  a many-body operator $\hat{O}$ in the standard form; 
	(g) the fermionic swap gate;
	and
	(h) the parity gate.
	}\label{fig:ferm_TN_dictionary}
	\end{figure}

\begin{figure}[b]
	\centering
	\includegraphics[width=0.5\columnwidth]{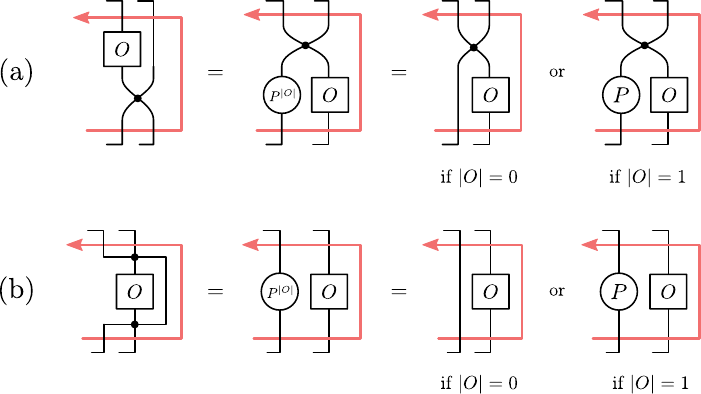}
	\caption{
	(a) Fermionic swap gates acting on one leg of tensor $\hat{O}$ can be exchanged with swap gates acting on the complementary set of legs of $\hat{O}$ if  $\hat{O}$ is even. If $\hat{O}$ is odd, an additional parity gate appears.
	(b) A line that crosses all legs of an operator $\hat{O}$ can be simplified according to the parity of the operator  $\hat{O}$.
	}\label{fig:ferm_identities}
	\end{figure}

Our diagrammatic representation of the fermionic tensor network approach is as follows:
\begin{enumerate}
	\item \textit{Fermionic ordering of $\mathbb{Z}_2$  graded tensor products} is represented by a single directed line in red passing through all elements of the super vector space (represented as \textit{open} black legs in Fig.~\ref{fig:ferm_TN_dictionary}).
	
	\item \textit{Kets (Bras) of the (dual) super vector space $V = V^0 \oplus V^1$} ($V^*$) are represented as open legs in black that point along (against) the direction of the arrow. 	
	
	\item \textit{Fermionic reordering} of $\ket{a_1}$ and $\ket{a_2}$ gives rise to a parity-dependent sign $(-1)^{|a_1||a_2|}$ which is represented as a crossing between two open legs, denoted as a black dot. 
\end{enumerate}
An example is the \textit{supertrace of a rank-2 operator $\hat{O}$}, which is written algebraically as
	\begin{align} \label{eq:supertrace}
 \mathrm{sTr} (\hat{O}) 
 :=
	\mathcal{C} \left( \sum_{a, b}  O_{ab} \ket{a} \otimesg \bra{b} \right) 
	\nonumber
 = \mathcal{C} \left( \sum_{a, b} O_{ab} (-1)^{|a| |b|} \bra{b} \otimesg \ket{a} \right) 
 = \sum_{a} (-1)^{|a|} \,  O_{aa}
= \mathrm{tr} ( P O).
	\end{align}	 
	Diagrammatically, we have
	\vspace{0.5mm}
	\begin{center}
	\includegraphics[width=0.45 \columnwidth]{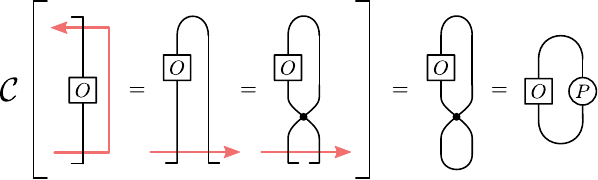}   
	\end{center}
	\vspace{0.5mm}
	where $P_{ij} =(-1)^{|i|} \delta_{ij}$ is the parity operator (see Fig.~\ref{fig:ferm_TN_dictionary}).
	Due to the additional parity operator appearing in the supertrace, the conventional fermionic trace operation of a rank-2 operator in the standard form is implemented as follows,
	\begin{equation}
\trf (\hat{O}) := \mathcal{C} \left( 
\sum_{a,b,c}
(-1)^{|c|} \ket{c} \otimesg \bra{c} \otimesg 
O_{ab} \ket{a} \otimes \bra{b}
\right) 	
=\trb (O) \;.
	\end{equation}
Lastly, we show two identities concerning the fermionic swap gates in Fig.~\ref{fig:ferm_identities}.


\section{2. Optimization of the quantum circuit $\tilde{U}$} 
In this section, we describe the optimization procedure, the goal of which is to variationally diagonalize the Hamiltonian  
\begin{align}\label{eq:hamiltonianSM}
    H = -J\sum_{\langle i,j \rangle} \sum_{\mu = \uparrow,\downarrow} (c^\dagger_{i,\mu}c_{j,\mu} + c_{j,\mu}^\dg c_{i,\mu}) +  V \sum_i  n_{i,\uparrow}n_{i,\downarrow} 
    + \sum_i W(i)(n_{i,\uparrow} + n_{i,\downarrow}),
\end{align}
where $W(i)=W(x,y)=\Delta\big(\cos(2\pi\beta_x x)+\cos(2\pi\beta_y y)\big)$ is the quasiperiodic potential. We use the letters $i,j$ to refer to a 2D coordinate: $i = (x,y) \in \mathbb{Z}^2\cap([1,N]\times[1,N])$.  We introduce the variational quantum circuit $\tilde U$, which can be viewed as a matrix whose columns are the approximate eigenstates of $H$, and which depends on the underlying variational parameters in a way that makes it unitary by construction. 

\subsection{2.1. Overview}
In order to optimize the approximate eigenstates in $\tilde U$, i.e. find the $\tilde U$ that approximately diagonalizes the Hamiltonian, we find approximate LIOMs $\tilde \tau_{i,\mu}^z (\tilde U) = \tilde U \sigma_{i,\mu}^z \tilde U^\dg$ which commute with the Hamiltonian $H$ as well as possible. To that end, we minimize a cost function~\cite{Wahl2017PRX,2DMBL} that penalizes the non-commutation of $\tilde \tau_{i,\mu}^z (\tilde U)$ and $H$:
\begin{align}
    f(\tilde U) &= \frac{1}{2N^24^{N^2}}\sum_i \sum_{\mu = \uparrow, \downarrow} \trf([H,\tilde \tau_{i,\mu}^z]^\dagger[H,\tilde \tau_{i,\mu}^z]) \notag\\
    &= \frac{1}{N^24^{N^2}}\sum_i \sum_{\mu = \uparrow,\downarrow} \big[ \trf(H^2(\tilde \tau_{\mu,i}^z)^2) - \trf(H\tilde \tau_{i,\mu}^z H \tilde \tau_{i,\mu}^z)\big], \label{eq:fom_SM} 
\end{align}
where to get to the second line, one simply expands out the commutators. In addition to the intuitive motivation, we also note that it is straightforward to prove that $f(\tilde U) = 0$ if and only if $\tilde U$ exactly diagonalizes the Hamiltonian (without assuming that $\tilde U$ is a quantum circuit or that the system is many body localized).  

In Eq.~\eqref{eq:fom_SM}, the sums over sites $i$ and spin $\mu$ produce $2N^2$ terms, and the dimension of the overall Fock space is $4^{N^2}$, so the factor of $1/(2N^24^{N^2})$ has been included to ensure that $f(\tilde U)$ becomes an ${\mathcal{O}(1)}$ number. The first term in the bracket reduces to the constant $\trf(H^2)$, since $(\tilde \tau_{i,\mu}^z)^2 = (\sigma_{i,\mu}^z)^2 = \mathbb{1}$. 
The cost function then effectively reduces to a sum of $- \trf(H\tilde \tau^z_{i,\mu} H \tilde \tau^z_{i,\mu})$ terms, each of which can be  calculated via a local tensor network contraction as explained in the following section.  

The optimization is possible for large values of $N$ since the computation time scales linearly with the system size due to the quantum circuit $\tilde U$ being built up of smaller unitaries acting on patches of $2 \times 2$ sites, cf. Fig~2a. 
Those unitaries are taken to be real and  particle number preserving, as the same applies to the Hamiltonian. (This is expected not to reduce the accuracy of the approximation significantly~\cite{Wahl2017PRX,2DMBL}.) The small unitaries are thus orthogonal block-diagonal matrices whose diagonal blocks can be parameterized as $u_\mr{block} = e^{A}$, where the real anti-symmetric matrix $A = -A^\top$ contains the (real) parameters to be optimized. 

For the Hamiltonian in Eq.~(1), we have $f(\mathbb{1}) = 4J$, regardless of the values of $V$ and $\Delta$.  
 We choose to initialize all unitaries (and consequently $\tilde U$) as identities and set $J=1$. Thus the cost function takes the initial value of $f=4$.
 After performing the optimization, we observe the cost function converging to $f\approx 2$ for small disorder strength ($\Delta\approx 1$), and to as small as $f\approx 0.1$ for large disorder strength ($\Delta > 20$).

Due to virtual memory constraints, we run the optimization by sweeping across the system and optimizing one small unitary at a time. We verified (using a spinless fermionic model) that the sweeping method results in no loss of accuracy compared to simultaneous updating of all parameters. 
The optimization was performed with the l-BFGS algorithm with automatic differentiation using the PyTorch library~\cite{Paszke2017}.

\subsection{2.2. Analytical evaluation of the local cost function}

\begin{figure*}[t]
    \centering
    \includegraphics[width=0.9 \textwidth]{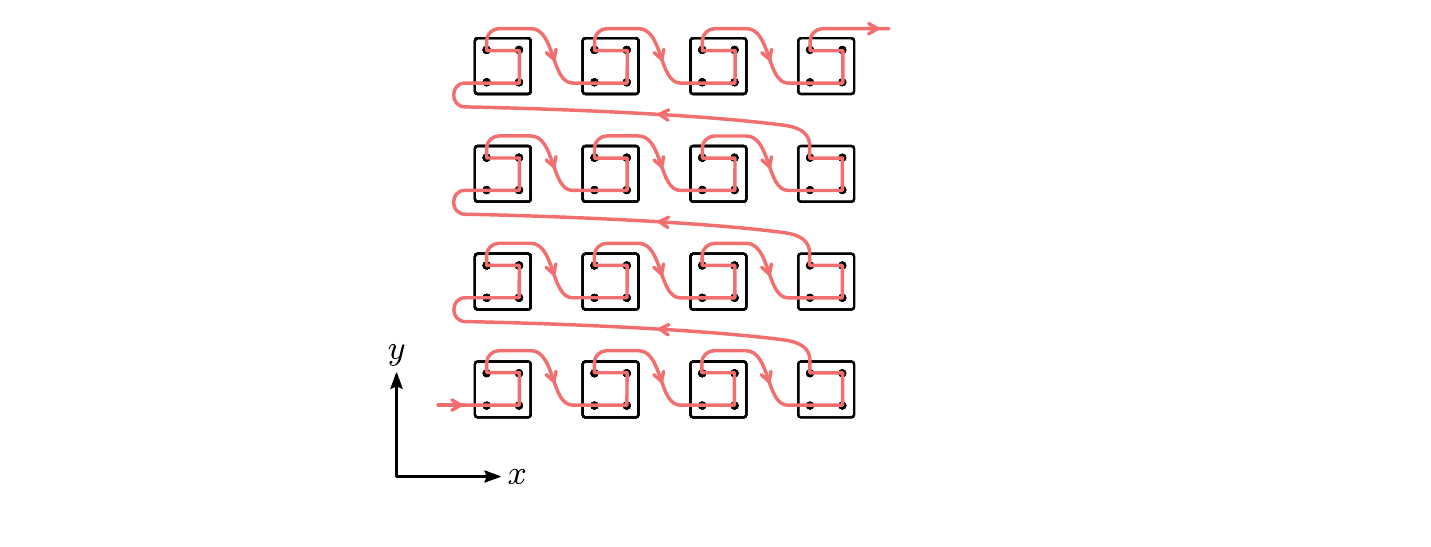}
    \caption{Chosen ordering of the fermionic operators. Each square contains four sites on which a red gate in Fig.~2 acts.}
    \label{fig:ordering}
\end{figure*}

\begin{figure*}[t]
\includegraphics[width=0.98 \textwidth  ]{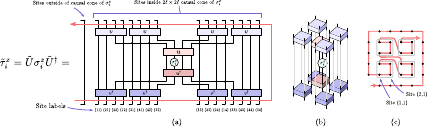}
    \caption{
    (a) Diagrammatic representation of $\tilde{\tau}_{i,\mu}^z= \tilde{U} \sigma_{i,\mu}^z \tilde{U}^\dagger$ as a 2D diagram. For simplicity, the index $\mu$ is suppressed in the diagrams. 
    The sites inside the causal cone of $ \sigma_{i,\mu}^z$ are illustrated explicitly, with the fermionic ordering specified in (c), while the ones outside are represented as a thick line. 
    (b) $\tilde{\tau}_{i,\mu}^z= \tilde{U} \sigma_{i,\mu}^z \tilde{U}^\dagger$ as a 3D diagram, where we have omitted the sites outside the causal cone of $ \sigma_{i,\mu}^z$ and the swap gates [explicitly shown in (a)] for simplicity. 
    (c) New fermionic ordering for the 32 sites inside of and surrounding the causal cone of $ \sigma_{i,\mu}^z$. The ordering of other sites in the system can be arbitrarily arranged once the gates in $\tilde{U}$ outside the causal cone of $\sigma_{i,\mu}^z$ have been canceled, as the identity lines can be dragged through the even operator $\tilde \tau_{i,z}^\mu$. Note that the constituting unitaries are site-dependent.
    } \label{fig:tau}
\end{figure*}

\begin{figure*}[t]
    \centering
    \includegraphics[width=0.98 \textwidth]{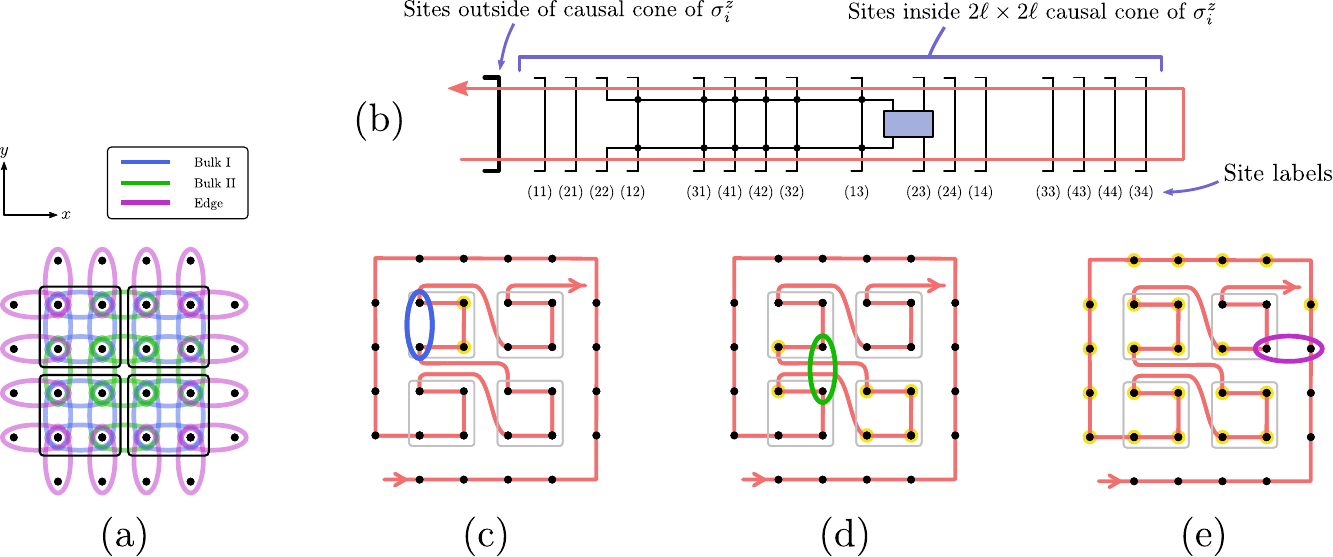}
    \caption{
    (a)
    Classification of three types of contributions to the Hamiltonian $H= \sum_{k} h_k$ for a given $\sigma^z_{i,\mu}$: 
    (i) Bulk-I terms (blue), which reside within $2 \times 2$ plaquettes (black squares); 
    (ii) bulk-II terms (green) which act across two plaquettes; and (iii) edge terms (purple) which straddle across the boundary of the causal cone. 
    (b) As an example, we illustrate the fermionic tensor network of a bulk-II type $h_k$ that acts on site $(2,2)$ and $(2,3)$. The coordinates within the causal cone are defined according to Fig.~\ref{fig:tau}c. The location of this term is illustrated in (d).
    (c) A bulk-I type $h_k$ (blue) can give rise to JW string (yellow) within a plaquette.
    (d) A bulk-II type $h_k$ (green) can give rise to JW string (yellow) in multiple plaquettes. The fermionic tensor network of this term is illustrated in (b).
    (e) An edge type $h_k$ (purple) can give rise to JW string (yellow) in multiple plaquettes and at sites surrounding the causal cone. 
    }
    \label{fig:plaqAcrossEdge}
\end{figure*}

In this subsection, we show how to compute the cost function contribution, $-\trf(H \tilde \tau_{i,\mu}^z H \tilde \tau_{i,\mu}^z)$, by evaluating the fermionic contractions within the fermionic trace operation. For the representation of the fermionic operators as matrices, we choose the fermionic ordering shown in Fig.~\ref{fig:ordering}. 
The operator $\tilde \tau_{i,\mu}^z = \tU \sigma_{i,\mu}^z \tU^\dagger$ is a contraction of fermionic operators,  
and is nontrivial only over a $4\times 4$ region due to its causal cone structure, which arises from the fact that all unitaries in $\tU$ and their conjugates in $\tU^\dagger$ cancel out (even in the presence of fermionic swap gates), except a local set of unitaries, as illustrated in Fig.~\ref{fig:tau}.
When calculating $-\trf(H \tilde \tau_{i,\mu}^z H \tilde \tau_{i,\mu}^z)$, we consider $H$ as a sum of local terms, and note that those terms that lie entirely outside the $4\times 4 $ causal cone region give a constant contribution to the cost function, and can therefore be ignored.  
By only including the Hamiltonian terms inside the causal cone, $-\tr(H \tilde \tau_{i,\mu}^z H \tilde \tau_{i,\mu}^z)$ becomes a sum of local tensor network contractions.

\footnotetext{To clarify the relation between $h_k^{\mr{loc}}$ and $h_k$:  $h_k^{\mr{loc}}$ is a 2-site operator, i.e. a $16\times 16$ matrix.  $h_k\equiv \mathbb{1} \otimes \ldots \otimes \mathbb{1} \otimes h_k^{\mr{loc}}\otimes \mathbb{1}\otimes \ldots \otimes\mathbb{1}$ is an operator that 
	acts \textit{non-trivially} on two sites but is formally a $4^{N^2}\times4^{N^2}$ matrix.}

\begin{figure*}[t]
    \centering
    \includegraphics[width=0.98 \textwidth]{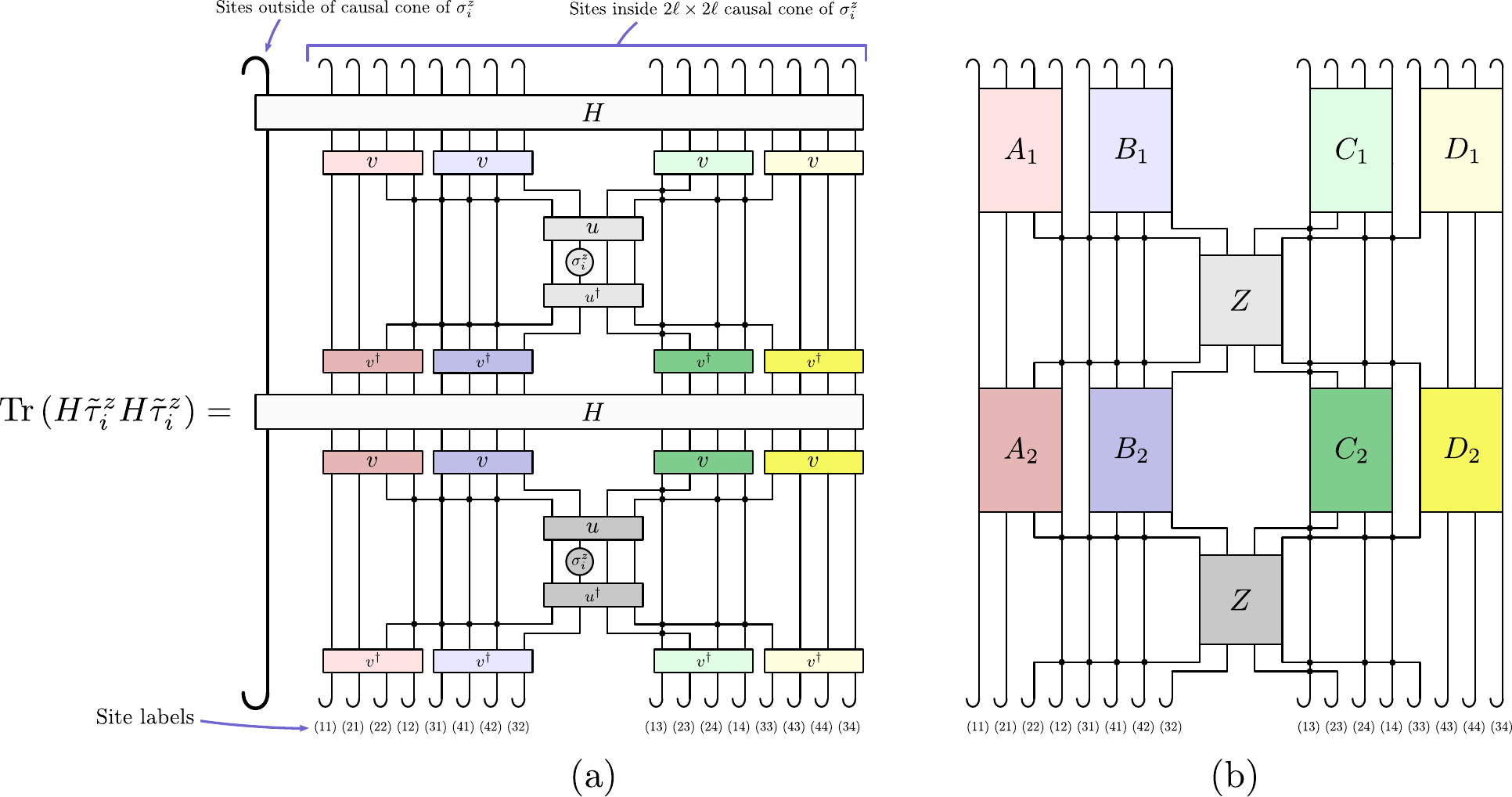}
    \caption{(a) Diagrammatic representation of the local contributions to the cost function. Additional subscripts of $u$, $v$, $\sigma_i^z$ and $\tilde \tau_i^z$ 
    have been suppressed for simplicity.
    By expanding the Hamiltonian $H= \sum_k h_k$ in terms of local contributions (see Figs.~\ref{fig:plaqAcrossEdge} and ~\ref{fig:h_decomp}), we can rewrite the cost function in (a) as a sum of tensor network contractions of the form in (b). The tensors with the same color in (a) are grouped into one tensor with the same color in (b) (after the decomposition of $H$).
    }
    \label{fig:fom_2d1d}
\end{figure*}

Specifically, we write the Hamiltonian $H = \sum_k h_k$ as a sum of 2-site operators over all pairs of nearest neighbors. The local action of the 2-site operators  $h_k$ is given by~\footnotemark[1] [ignoring Jordan-Wigner~(JW) strings of fermion parity operators on other sites; see next paragraph]
\begin{align}
    h_k^{\mr{loc}} = -J\left((c^\dag_\uparrow Z) \otimes c_\uparrow + (c^\dag_\downarrow Z) \otimes c_\downarrow +(Z  c_\uparrow) \otimes c^\dag_\uparrow + (Z c_\downarrow) \otimes c^\dag_\downarrow \right) + \frac{1}{2} \left[Vn_\uparrow n_\downarrow + W(x,y)(n_\uparrow + n_\downarrow) \right]\otimes\mathbb{1} \; ,
    \label{eq:h_k} 
\end{align}
where $Z = \text{diag}(1,-1,-1,1)$ is the local parity operator, and 
$c^\dag_\mu$ and $c_\mu$ are creation and annihilation operators acting on a single site (i.e. $4\times 4$ matrices). $n_\mu = c_\mu^\dg c_\mu$ 
and $W(x,y) = \Delta[\cos(2\pi(\beta_x x + \delta x))+\cos(2\pi(\beta_y  y + \delta y ))]$, where $(x,y)$ is the coordinate of one of the two sites that $h_k$ acts on (we will use the convention that it is always the coordinate of the left or lower site, when $h_k$ lies horizontally or vertically, respectively). 
We thus have $-\Tr(H \tilde \tau_{i,\mu}^z H \tilde \tau_{i,\mu}^z) = -\sum_{k,l} \Tr(h_k \tilde \tau_{i,\mu}^z h_l \tilde \tau_{i,\mu}^z)$.  
Now we partition the $4\times 4$ causal cone region (where $\tilde \tau_{i,\mu}^z$ is non-trivially supported) into four $2 \times 2$ plaquettes as in the black squares in Fig.~\ref{fig:plaqAcrossEdge}a.  
The $h_k$'s that act non-trivially on the $4\times 4$ causal cone region can be classified as (Fig.~\ref{fig:plaqAcrossEdge}a):
(i) Bulk-I terms, which reside within one of the plaquettes;
   (ii) bulk-II terms, which act across two plaquettes; 
   and (iii) edge terms, which straddle across the boundary of the causal cone of $\tilde \tau^z_{i,\mu}$. 
If both $h_k$ and $h_l$ are  bulk-I terms, i.e. they lie within plaquettes, $h_k \tilde \tau_{i,\mu}^z h_l \tilde \tau_{i,\mu}^z$ reduces to an expression of the form of Fig.~\ref{fig:fom_2d1d}b, which is relatively easy to contract (see the description of the contraction algorithm below).  If either $h_k$ or $h_l$ is a bulk-II or edge term, i.e. $h_{k/l}$ lies across plaquettes or the edge, then $h_k$ or $h_l$ is decomposed using Eq.~\eqref{eq:h_k} into a sum of products of on-site operators, so we end up expressing  $h_k \tilde \tau_{i,\mu}^z h_l \tilde \tau_{i,\mu}^z$ as a sum of multiple expressions of the form Fig.~\ref{fig:fom_2d1d}b.

\begin{figure*}[t]
\centering
\includegraphics[width=0.98 \textwidth  ]{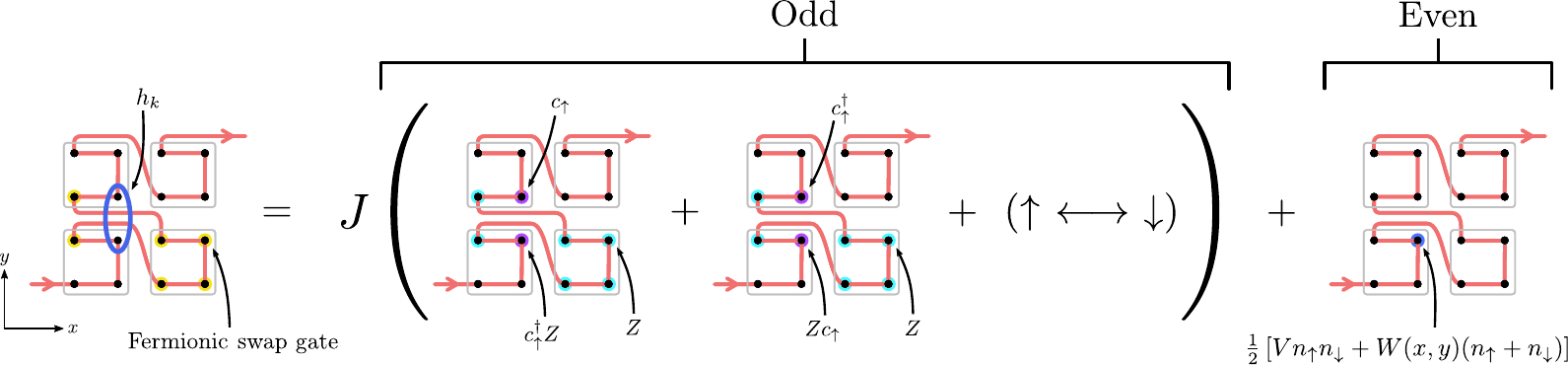}
    \caption{A graphical depiction of Eq.~\eqref{eq:h_k}, for an example of a Bulk-II type $h_k$ operator. 
    On the left, we illustrate $h_k$ in the notation of Fig.~\ref{fig:plaqAcrossEdge} (sites outside of the $4 \times 4$ causal cone are not shown). 
    On the right, we illustrate a sum of five tensor products. 
    } \label{fig:h_decomp}
\end{figure*}

\begin{figure*}
\centering
\includegraphics[width=0.99 \textwidth  ]{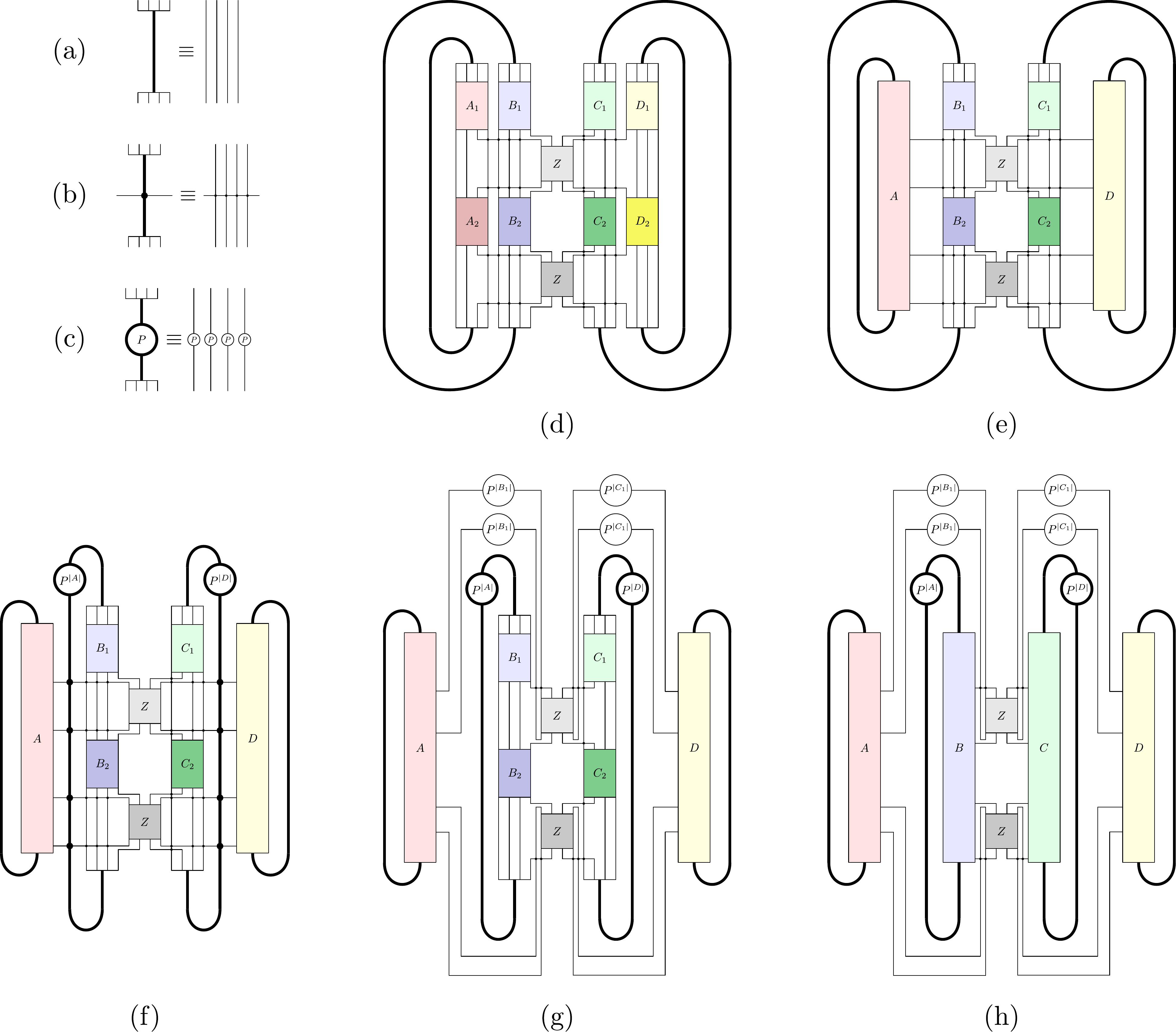}
    \caption{
    (a) Definition of a thick line, which represents a collection of four parallel thin lines.  
    (b) A fermionic swap gate or (c) a parity gate acting on a thick line represents fermionic swap gates or parity gates applied separately on the four parallel thin lines.  
    (d) A diagrammatic representation of contributions to the cost function after the evaluation of the fermionic contraction (equivalent to Fig.~\ref{fig:fom_2d1d}).
    (e) Definition of tensor $A$ ($D$) as a combination of the $A_1$ and $A_2$ tensors in red ($D_1$ and $D_2$ in yellow).
    (f) Dragging the outermost thick lines in (e) towards the center of the diagram, which gives rise to fermionic swap gates and parity operators according to the identities in Fig.~\ref{fig:ferm_identities}.
    (g) Dragging the four thin lines connected to tensor $A$ (and $D$) to the top and the bottom of the diagram. Again, fermionic swap gates and parity operators appear according to the identities in Fig.~\ref{fig:ferm_identities}.
    (h) The final simplified diagram is obtained by defining tensor $B$ ($C$) as a combination of the $B_1$ and $B_2$ tensors in blue ($C_1$ and $C_2$ in green).
    } \label{fig:trace1}
\end{figure*}

As mentioned above, it is not strictly true that the 2-site operators $h_k$ can be locally written as in Eq.~\eqref{eq:h_k}, due to the presence of JW strings.  
Specifically, if the 2-site operator $h_k$ lies along the fermionic ordering in Fig.~\ref{fig:tau}c, then the previous paragraphs apply directly.  
However, if $h_k$ does not lie along the fermionic ordering, then the fermionic contractions give rise to a series of fermionic swap gates, which we call a JW string, located along the fermionic ordering and in between the creation and annihilation operators of $h_k$.  
As an example, we consider a bulk-II term $h_k$ illustrated in Fig.~\ref{fig:plaqAcrossEdge}d, which can be expressed in the tensor network notation as Fig.~\ref{fig:plaqAcrossEdge}b.  This decomposition and resulting JW strings are shown in Fig.~\ref{fig:h_decomp}.  
Generally, a JW string appears for bulk-I term $h_k$ if $h_k$ acts within a plaquette on the bottom left and top left site, and for all bulk-II and edge terms.

In the paragraphs above, we have evaluated the fermionic contraction in the local cost function $-\Tr(H \tilde \tau_{i,\mu}^z H \tilde \tau_{i,\mu}^z)$.
We showed that the cost function involves calculating $h_k \tilde \tau_{i,\mu}^z h_l \tilde \tau_{i,\mu}^z$ terms and summing over all the $h_k$ and $h_l$ operators that intersect the $4\times 4$ causal cone of $\sigma_{i,\mu}^z$. 
By decomposing $h_k \tilde \tau_{i,\mu}^z h_l \tilde \tau_{i,\mu}^z$ whenever necessary, we can write the local cost function in the form depicted in Fig.~\ref{fig:fom_2d1d}b:  
We further simplify Fig.~\ref{fig:fom_2d1d}a using tensor network techniques.
For a given $h_k$ and $h_l$, we combine pairs of unitaries ($u^\dg$, $u$) and ($v^\dg$, $v$) with the intervening Hamiltonian terms in Fig.~\ref{fig:fom_2d1d}a to obtain the diagram in Fig.~\ref{fig:fom_2d1d}b.

Next, we simplify Fig.~\ref{fig:fom_2d1d}b in Fig.~\ref{fig:trace1} by contracting together tensors and dragging lines across tensors using the identities described in Fig.~\ref{fig:ferm_identities}.
In Fig.~\ref{fig:trace1}, we adopt the convention of thick lines which represent four parallel thin lines as described in Fig.~\ref{fig:trace1}a-c. 
We use the intermediate steps shown in Fig.~\ref{fig:trace1}d-g to arrive at the final simplified diagram in Fig.~\ref{fig:trace1}h.  

In short, the cost function $-\trf(H \tilde \tau_{i,\mu}^z H \tilde \tau_{i,\mu}^z)$ can be evaluated as Fig.~\ref{fig:fom_2d1d}a. After expanding the Hamiltonian in terms of local contributions given in Fig.~\ref{fig:plaqAcrossEdge} using the expansion in Fig.~\ref{fig:h_decomp}, the fermionic contractions within the trace can be performed, and the cost function can be expressed as a sum of tensor networks of the form shown in Fig.~\ref{fig:trace1}h.

\subsection{2.3. Numerical computation of the local cost function}
In practice, it is not the most efficient way to loop over $h_k$ and $h_l$. Some speedup can be achieved by organizing the sums in specific ways.  Additionally, some subtleties arise when $h_k$ and $h_l$ lie on the edge of the causal cone region. 
In this section, we discuss the algorithm used to numerically calculate the local cost function as a sum over tensor networks of the type shown in Fig.~\ref{fig:trace1}h. 

First, we make a parenthetical comment about the dimensionality of operators.  In the below, we will refer to operators that act only on a local region.  In the interest of making the equations consistent in the simplest way possible, let us adopt the convention that operators are defined on the full Fock space.  So for example, $h_k$ acts non-trivially on 2 sites in the $4^{N^2}$-dimensional Fock space.

Our aim is to compute $\trf(H \tilde \tau_{i,\mu}^z H \tilde \tau_{i,\mu}^z)$. Recall that only Hamiltonian terms that have non-trivial support in the $\tilde \tau_{i,\mu}^z$ causal cone need to be considered. 
For a given causal cone, we decompose the Hamiltonian as $H = \sum_{k \in \text{ outside}} h_k + \sum_{k \in \text{ causal cone}} h_k \equiv \Hocc + \Hcc$.  
Within the sum over terms in the causal cone, $h_k$ terms may lie within a plaquette (Bulk I), across two plaquettes but still in the bulk (Bulk II), 
or on the edge, as depicted in Fig.~\ref{fig:plaqAcrossEdge}.  
Let us write
\begin{align}
    \Hcc =
    \sum_{\substack{k\,\in \\ \text{Bulk I}}} h_k + 
    \sum_{\substack{k\,\in \\ \text{Bulk II}}}h_k +  
   \sum_{\substack{k\,\in \\ \text{Edge}}} h_k \; .
\end{align}
Furthermore, we define
\begin{align}
    H_\Box \equiv
      \sum_{\substack{k\,\in \\ \text{Bulk I}}} h_k 
    +  
    \sum_{\substack{k\,\in \\ \text{Edge}}} h_k'  \; ,
\end{align}
where $h'_k$ is the operator acting on the edge with the outside leg traced out,
\begin{align}
    \includegraphics[width=0.15 \textwidth]{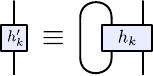}
\end{align}
where both operators have to be thought of as tensor products with $\mathbb{1}$ on all other sites. 
In other words, $H_\Box$ is defined to be $\Hcc$ but without the Bulk II terms, and with the edge legs traced out (so it is an operator acting non-trivially on only 16 sites instead of 32 sites).
We can now write the local cost function in a form that involves much fewer tensor network contractions, 
\begin{align}\label{eq:fABCD}
   \Tr(\Hcc \tilde \tau_{i,\mu}^z \Hcc \tilde \tau_{i,\mu}^z) \; = \;
   &
   \underbrace{\Tr(H_\Box \tilde \tau_{i,\mu}^z H_\Box \tilde \tau_{i,\mu}^z)}_{\mbox{\large $\fA$}} \;+\; \underbrace{2\sum_{\substack{k \in \\ \text{Bulk II}}} \Tr(H_\Box \tilde \tau_{i,\mu}^z h_k \tilde \tau_{i,\mu}^z)}_{\mbox{\large $\fB$}} \notag 
   \\ 
   &
   + \;
   \underbrace{\sum_{\substack{k \in \\ 
   \text{Bulk II}}}\sum_{\substack{l \in \\ \text{Bulk II}}} \Tr(h_k \tilde \tau_{i,\mu}^z h_l \tilde \tau_{i,\mu}^z)}_{\mbox{\large $\fC$}}  \;\,
   \;
   +  \;
   \underbrace{
    \sum_{\substack{k\,\in \\ \text{Edge}}} 
    \Big[ \Tr(h_k \tilde \tau_{i,\mu}^z h_k \tilde \tau_{i,\mu}^z) - \Tr(h'_k \tilde \tau_{i,\mu}^z h'_k \tilde \tau_{i,\mu}^z)    \Big]}_{\mbox{\large $\fD$}}
    \;
    .
\end{align}
  The four parts labeled $\fA$, $\fB$, $\fC$, and $\fD$ can now be calculated separately, with $\fC$ being the most computationally expensive term to calculate.

  Note that the right hand side of Eq.~\eqref{eq:fABCD}, with the exception of the first term in $\fD$, only contains operators that act non-trivially within the $4\times4$ region.  Thus we can take the trace with the above described contraction algorithm, and then normalize by a factor of $1/4^{16}$. (For the first term in $\fD$, we need to normalize by $1/4^{17}$, as one of the tensors $A$, $B$, $C$, $D$ obtains one additional leg.)
The bulk-II $h_k$ operators as well as the edge $h_k$ operators in the first term of $\fD$, need to be decomposed into a sum of tensor products, as in Eq.~\eqref{eq:h_k}.  
As a result, the sum over the bulk-II operators is a sum over 40 terms: 8 bulk-II terms $h_k$, multiplied by 5 contributions per $h_k$.  
Thus, we can see that the calculation of $\fB$ requires 40 trace calculations, and the calculation of $\fC$ manifestly requires 1600 trace calculations (note that, for the latter case, the number of terms can be halved due to symmetry).

\section{3. Calculation of entanglement entropy}

Here we describe the methods used to calculate the transition point $\Deltac$ from the optimized quantum circuit $\tilde U$.  Specifically, we describe the computation of the entanglement entropy and its standard deviation (which gives Fig.~1) for a certain filling fraction $\nfill$. 

\begin{figure*}[t]
\centering
\includegraphics[width=0.95 \textwidth  ]{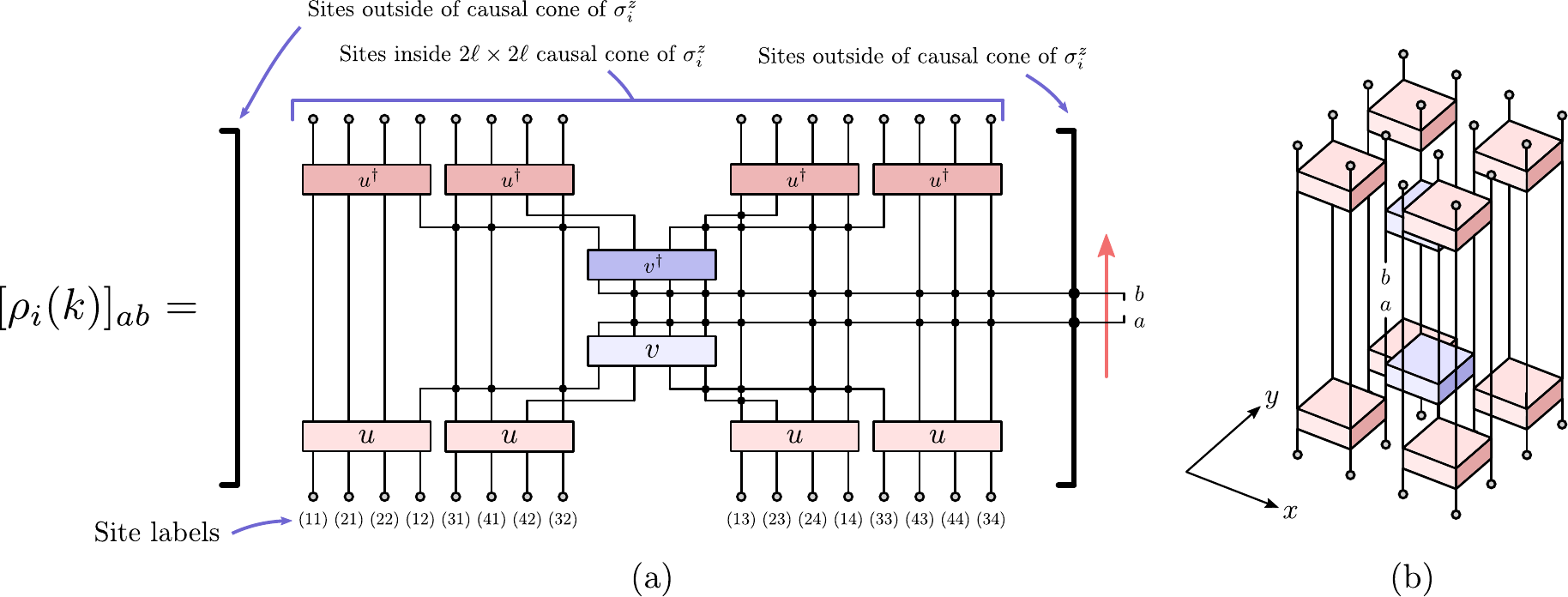}
    \caption{
    (a) Diagrammatic representation of $[\rho_i(k)]_{ab} = \trf_{\overline{i}}\left[|\psi(k)\rangle\langle\psi(k)|\right]_{ab}$ as a 2D diagram. 
    The sites inside the causal cone of site $i$ are illustrated explicitly, while the ones outside are represented as a thick lines. 
    (b) $[\rho_i(k)]_{ab} = \trf_{\overline{i}}\left[|\psi(k)\rangle\langle\psi(k)|\right]_{ab}$  as a 3D diagram, where we have omitted the sites outside the causal cone of $ \sigma_{i,\mu}^z$ and the swap gates (explicitly shown in (a)) for simplicity. Note that due to the even parity of the unitaries, $[\rho_i(k)]_{ab}$ is non-vanishing only for $|a| + |b| \mod 2 = 0$. 
    }
    \label{fig:rhoDefinition}
    \label{fig:rho2D}
\end{figure*}

For each site $i$ we first calculate the on-site entanglement entropies averaged over 1000 eigenstates. Our main quantity of interest is the standard deviation of this average entropy with respect to disorder realizations averaged over sites $i$~\cite{2DMBL},
\begin{align}
    \sigma_S(\Delta) = \frac{1}{N^2}\sum_i\sigma_{S,i}(\Delta) \; ,
\end{align}
where $N$ is the linear system size, and $\sigma_{S,i}(\Delta)$ is the standard deviation of the on-site entropies at site $i$,
\begin{align}
    \sigma_{S,i}(\Delta) = \sqrt{ \left\langle S_i^2(\Delta) \right\rangle- \left\langle S_i(\Delta) \right\rangle^2} \; .
\end{align}
Here $\langle \cdot \rangle$ means an average over all disorder realizations at disorder strength $\Delta$ (we used 10 disorder realizations at each $\Delta$), and $S_i(\Delta)$ is the on-site entanglement entropy at site $i$ averaged over $N_\mr{states} = 1000$ approximate eigenstates $k$,
\begin{align}
    S_i(\Delta) = -\frac{1}{N_\text{states}}\sum_{k=1}^{N_\text{states}}\tr(\rho_i(k) \ln\rho_i(k)) \;.
\end{align}
The density matrix $\rho_i(k)$ is given by 
\begin{align}
    \rho_i(k) = \trf_{\overline{i}}\left[|\psi(k)\rangle\langle\psi(k)|\right] \;,
\end{align}
where the partial trace is taken over all sites other than $i$, and $|\psi(k)\rangle$ is a randomly chosen approximate eigenstate from $\tilde U$ obtained by fixing the indices corresponding to the lower legs of Fig.~2a. 
The graphical representation of this equation is depicted in Fig.~\ref{fig:rhoDefinition}.  Due to the quantum circuit structure, the calculation of $\rho_i(k)$ becomes local. 

We can also calculate $\sigma_S(\Delta,\nfill)$, the standard deviation of the entropy at some given filling fraction $\nfill$. Recall that all (approximate) eigenstates have well-defined particle number, as $\tilde U $ is particle-number conserving.  The procedure is exactly as described above for calculating $\sigma_S(\Delta)$, except we constrain the random states $|\psi(k)\rangle$ to be of filling fraction $\nfill$ by choosing the eigenstates $k$ such that they correspond to an l-bit configuration with overall particle number $\nfill N^2$.  Also, when calculating the filling fraction dependent entropy $S_i(\Delta,\nu)$, we average over $10^4$ instead of 1000 eigenstates, due to the increased sensitivity to fluctuations in this case.

To obtain the $\nu$-dependent phase diagram as shown in Fig.~1, we compute $\sigma_S(\Delta,\nu)$ for the 29 filling fractions $\nu \in [0.3,1.7]$ in steps of 0.05.  
That is, we performed the procedure described above, at each stage constraining the random states $|\psi(k)\rangle$ to have particle numbers $N^2 \nu \in [30, 170]$ in steps of 5.
These plots, with a third-order polynomial fit, are shown in Fig. 3b 
We located the maxima of these curves to find the filling fraction-resolved transition points $\Delta_c(\nu)$, which we plot (together with a fitted fourth order polynomial) to form the filling fraction-dependent phase diagram of Fig.~1. 
%
%

\section{4. Calculation of localization lengths}

In this section we discuss the numerically efficient calculation of the localization lengths using Eq.~(3). 
Assuming the operator $\mathcal{O}$ is localized on some site $j$, we would expect $w_i(\mathcal{O})$ to to scale as $\exp(-\text{dist}(i,j)/\xi_L)$, where $\xi_L$ is the localization length. 
Given an optimized $\tilde U$, we can use the weight functions of the l-bit operators to discern the average localization length of the system.  As we have two l-bit operators per site, we define the weight $w_{ij} \equiv w_i(\tilde \tau_{j,\uparrow}^z) + w_i(\tilde \tau_{j,\downarrow}^z)$.

For the weight function of the l-bit operators, we have 
\begin{align}
    w_i(\tilde \tau_{j,\mu}^z) = {\sum_{\beta,\gamma}}' \Tr\left[\left(\tilde \tau_{j,\mu}^z - \hat\sigma_{i,\uparrow}^\beta \hat \sigma^\gamma_{i,\downarrow}\tilde \tau_{j,\mu}^z \hat \sigma^\gamma_{i,\downarrow} \hat\sigma^\beta_{i,\uparrow}\right)^2\right] = 2\,{\sum_{\beta,\gamma}}' \Tr\left[\mathbb{1} - \tilde\tau_{j,\mu}^z \hat\sigma_{i,\uparrow}^\beta \hat \sigma^\gamma_{i,\downarrow}\tilde \tau_{j,\mu}^z \hat \sigma^\gamma_{i,\downarrow} \hat\sigma^\beta_{i,\uparrow}\right],
\end{align}
where $\Tr\big(\tilde \tau_{j,\mu}^z \hat\sigma_{i,\uparrow}^\beta \hat \sigma^\gamma_{i,\downarrow}\tilde \tau_{j,\mu}^z \hat \sigma^\gamma_{i,\downarrow} \hat\sigma^\beta_{i,\uparrow}\big)$ can be calculated efficiently by the same methods used to calculate $\Tr\big(\tilde\tau_{j,\mu}^z h_k \tilde \tau_{j,\mu}^z h_l)$ for the cost function.  As explained in the main text, the sum $\sum'$ is a sum over all possible $\beta, \gamma \in \{0,x,y,z\}$ except $\beta = \gamma = 0$, i.e. there are 15 terms in this sum. 

Note that $w_{ij}$ will be  zero if the site $i$ does not lie in the causal cone of $\tilde \tau_{j,\mu}^z$. Thus for l-bit operators on a given site $j$, we can calculate $w_{ij}$ for the sixteen sites $i$ that lie in the causal cone.  Due to the geometry, the distance between sites $i$ and $j$ will be either $0,1,\sqrt{2},2,\sqrt{5},$ or $2\sqrt{2}$.  For a given site $j$ and the 16 sites $i$, we plot $w_{ij}$ as a function of $\text{dist}(i,j)$ and fit the data to an exponential curve $Ae^{-\text{dist}(i,j)/\xi_{L,j}}$ to find the localization length $\xi_{L,j}$.  We first average the fitted $\xi_{L,j}$ over all sites $j$, and then average over all disorder realizations at a given $\Delta$, to obtain $\xi_L(\Delta)$.  

For the Anderson localization lengths, we diagonalized the Hamiltonian of the equivalent single particle model, a tight-binding model on a square lattice with
\begin{align}
    H = -J\sum_{\langle i,j \rangle} \big(\ket{i}\bra{j}+\ket{j}\bra{i}\big) + \sum_i W(i) \ket{i}\bra{i},
\end{align}
where $W(i)$ is the same as in the many-body model.  For a particular eigenstate $\ket{\psi}$, and defining $\psi(i)=\braket{i|\psi}$, we found the localization length $\xi_L(\psi)$ by performing curve fitting on $|\psi(i)|^2\propto\exp[-\text{dist}(i,i_0)/\xi_L(\psi)]$, where $i_0=\text{argmax}(|\psi(i)|^2)$ is the site on which $\ket{\psi}$ is localized.  We then average over eigenstates and disorder realizations  to obtain $\xi_L(\Delta)$.

\section{5. Error analysis}

Here we describe how the error bars in Fig.~3 were determined. These indicate $\pm 1$ standard error of the standard deviation of the eigenstate-averaged on-site entanglement entropy with respect to disorder realizations, averaged over site positions. Hence, there will be statistical errors due to (i) the approximate eigenstate average ($M = 1000$ samples), (ii) the finite number of disorder realizations, (iii) the site average. We consider each of these contributions in turn:

First, if $S_i(m,r)$ is the entropy at site $i$ of the $m$-th randomly sampled approximate eigenstate of the Hamiltonian for disorder realization $r$, we have the eigenstate average $\overline{S_i}(r) = \frac{1}{M} \sum_{m=1}^M S_i(m,r)$. Its standard error is 
\begin{align}
\Delta \overline{S_i}(r) \approx \frac{1}{\sqrt{M}}\sqrt{\frac{1}{M}\sum_{m=1}^M\left[S_i(m,r) - \overline{S_i}(r) \right]^2}.
\end{align}
Second, the standard deviation of $\overline{S_i}(r)$ with respect to disorder realizations $r$ is
\begin{align}
\sigma_i = \sqrt{\frac{1}{R} \sum_{r=1}^R\left[\overline{S_i}(r) - \langle \overline{S_i}\rangle\right]^2},
\label{eq:Delta_S}
\end{align}
where $\langle \overline{S_i}\rangle = \frac{1}{R} \sum_{r=1}^R\overline{S_i}(r)$, and $R = 10$ is the number of disorder realizations. $\sigma_i$ will have two contributions to its standard error, one from the finite number of disorder samples, and one due to the error $\Delta \overline{S_i}(r)$ of $\overline{S_i}(r)$,
\begin{align}
\Delta \sigma_i \approx \sqrt{\frac{1}{4R}\left(\frac{\mu_i^{(4)}}{\sigma_i^2} - \frac{R-3}{R-1}\sigma_i^2\right)} + \sqrt{\sum_{r=1}^R \left( \frac{\partial \sigma_i}{\partial \overline{S_i}(r)}\right)^2 [\Delta S_i(r)]^2}.
\end{align}
The first term is an estimate for the standard error of the standard deviation~\cite{Rao1978}, which becomes exact in the limit of large sample size $R \rightarrow \infty$. $\mu_i^{(4)}$ denotes the fourth order moment, which based on our sample data can be estimated as $\mu_i^{(4)} \approx \frac{1}{R} \sum_{r=1}^R \left(\overline S_i(r) - \langle \overline S_i \rangle \right)^4$. The second term corresponds to error propagation of approximately uncorrelated random variables and can be evaluated by using Eq.~\eqref{eq:Delta_S} and $\frac{\partial \sigma_i}{\partial \overline{S_i}(r)} = \frac{R-1}{R^2\sigma_i}\left(\overline{S_i}(r) - \langle \overline{S_i}\rangle\right)$. On average, the second term contributes only about $4\%$ of $\Delta \sigma_i$ and was hence neglected in our numerical calculations for simplicity.

The site-average of the entanglement entropy fluctuatio6n is given by $\sigma_S = \frac{1}{N^2} \sum_{i=1}^{N^2} \sigma_i$. Hence, its standard error is
\begin{align}
\Delta \sigma_S \approx 
\frac{1}{N^2} \sqrt{\sum_{i=1}^{N^2} \Delta \sigma_i^2}, 
\end{align} 
corresponding to error propagation of approximately uncorrelated random variables.

\begin{figure*}
	\centering
	\includegraphics[width=0.46\textwidth ]{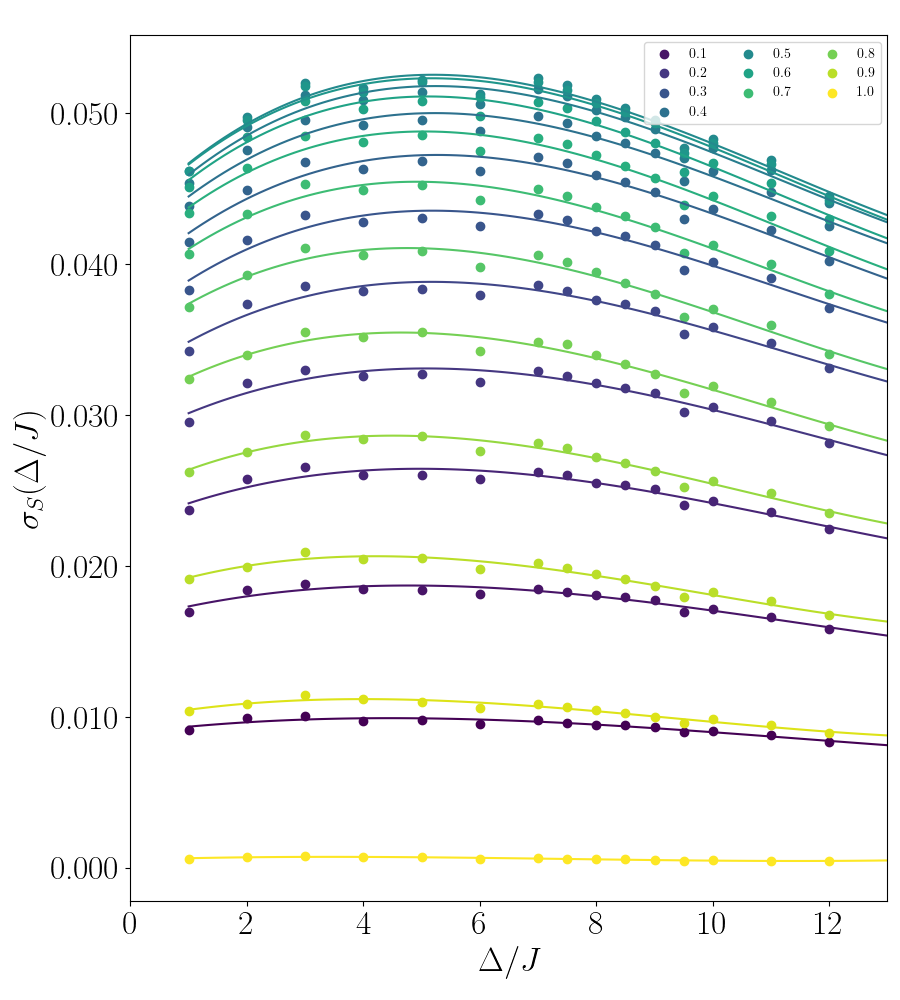} \ \ 
	\includegraphics[width=0.48 \textwidth ]{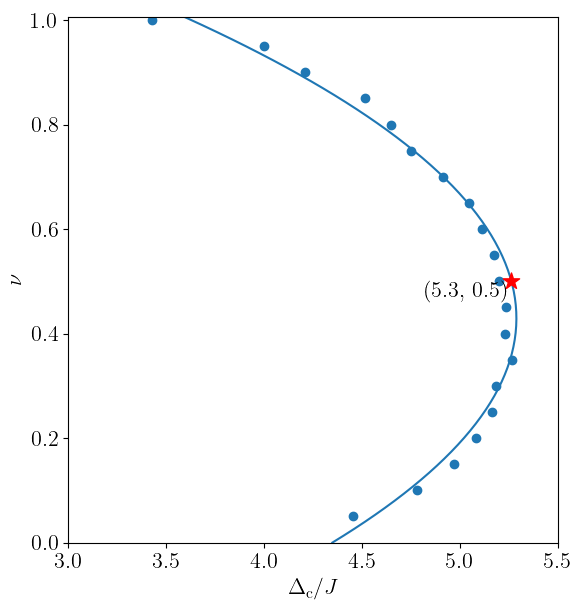}
	\caption{Left: Entanglement entropy fluctuation for the truncated model with on-site Hilbert space dimension $d = 3$ resolved with respect to individual filling fractions (indicated in the legend) with cubic fits. We used 5000 eigenstates to sample for each filling fraction. Right: Mobility edge extracted from the corresponding maxima. The filling fraction ranges in the regime $[0,1]$, as only up to one fermion per site (spin-up or spin-down) is allowed. Note that at $\nu = 1$ all eigenstates are product states, because the hopping term in the Hamiltonian becomes irrelevant. This is the effect leading to localization for large $\nu$, rather than the disorder.
	}
	\label{fig:entanglement_fluctuation}
\end{figure*}

\section{6. Numerical results for on-site Hilbert space dimension $d = 3$}

Here we present our numerical results for the model (and tensor network) truncated to local dimension $d = 3$. This will give an indication of the systematic error inherent to our approximation due to the finite entanglement allowed in our tensor network ansatz. In order to study the effect of the entanglement truncation, we simulated a similar model as Eq. (1) but where double occupancies are excluded.
In this truncated model, the on-site Hilbert space dimension is $d=3$, with basis states $(0,\uparrow,\downarrow)$ instead of $(0,\uparrow,\downarrow,\uparrow\downarrow)$ and the Hamiltonian is the original Eq. (1) but with the term $V \sum_i n_{i,\uparrow} n_{i,\downarrow}$ ignored. Note that $d$ equals the local dimension of the gates and thus the bond dimension of the tensor network.

In Fig.~\ref{fig:entanglement_fluctuation} we show the obtained entanglement entropy fluctuations for the same 10 disorder realisations as in the main body of the paper and the extracted mobility edge. We find a phase transition point at around $\Delta_c^{d = 3}(0.5) = 5.3J$. One can mitigate the effect of the exclusion of double occupancies to some extent by realizing that it implies an effective reduction of the contribution of the hopping term in Eq.~(1): If one considers any fermion in the system, on average its nearest neighboring sites will be occupied by two fermions, one with the same spin, and one with opposite spin. Due to the exclusion of double occupancies, the considered fermion cannot hop to the site occupied by a fermion with opposite spin, effectively reducing the hopping by a factor of 3/4, which also lowers the disorder strength required to localize the system. After accounting for this reduction, we obtain a better estimate for the phase transition point only due to entanglement truncation in the ansatz of ${\Delta_c'}^{d=3}(0.5) \approx 3/4 \cdot \Delta_c^{d = 3}(0.5) \approx 7 J$, close to the value at $d = 4$, $\Delta_c^{d = 4}(0.5) = 8.3 \pm 0.4J$. A comparison of $d = 3$ and $d = 4$ in the limit $\nu \rightarrow 0$ further corroborates our approach: In that limit, the above effect (and double occupancies) are neglegible, and indeed we obtain $\Delta_c^{d = 3}(\nu \rightarrow 0) = 4.3J$ according to Fig.~\ref{fig:entanglement_fluctuation}b and $\Delta_c^{d=4}(\nu \rightarrow 0) \approx 3J$ according to Fig.~1, in similar agreement as ${\Delta_c'}^{d=3}(0.5)$ and $\Delta_c^{d=4}(0.5)$. Crucially, also in this case we observe a drift to the real value, which is the exactly known non-interacting transition point of $\Delta_c^\mr{non-i}(\nu \rightarrow 0) = 2J$. 
Note that in the opposite limit of $\nu \rightarrow 1$, for $d = 3$ the hopping term in Eq.~(1) becomes irrelevant, as the fermions are unable to hop anywhere. The eigenstates are thus given by product states, and we expect $\Delta_c \rightarrow 0$. Our data does not fully reproduce this trend, although the upper branch of Fig.~\ref{fig:entanglement_fluctuation}b is substantially bent to the left.

\section{7. Distributions of entanglement entropies near the transition}

\begin{figure*}
\centering
\includegraphics[width=0.32\textwidth  ]{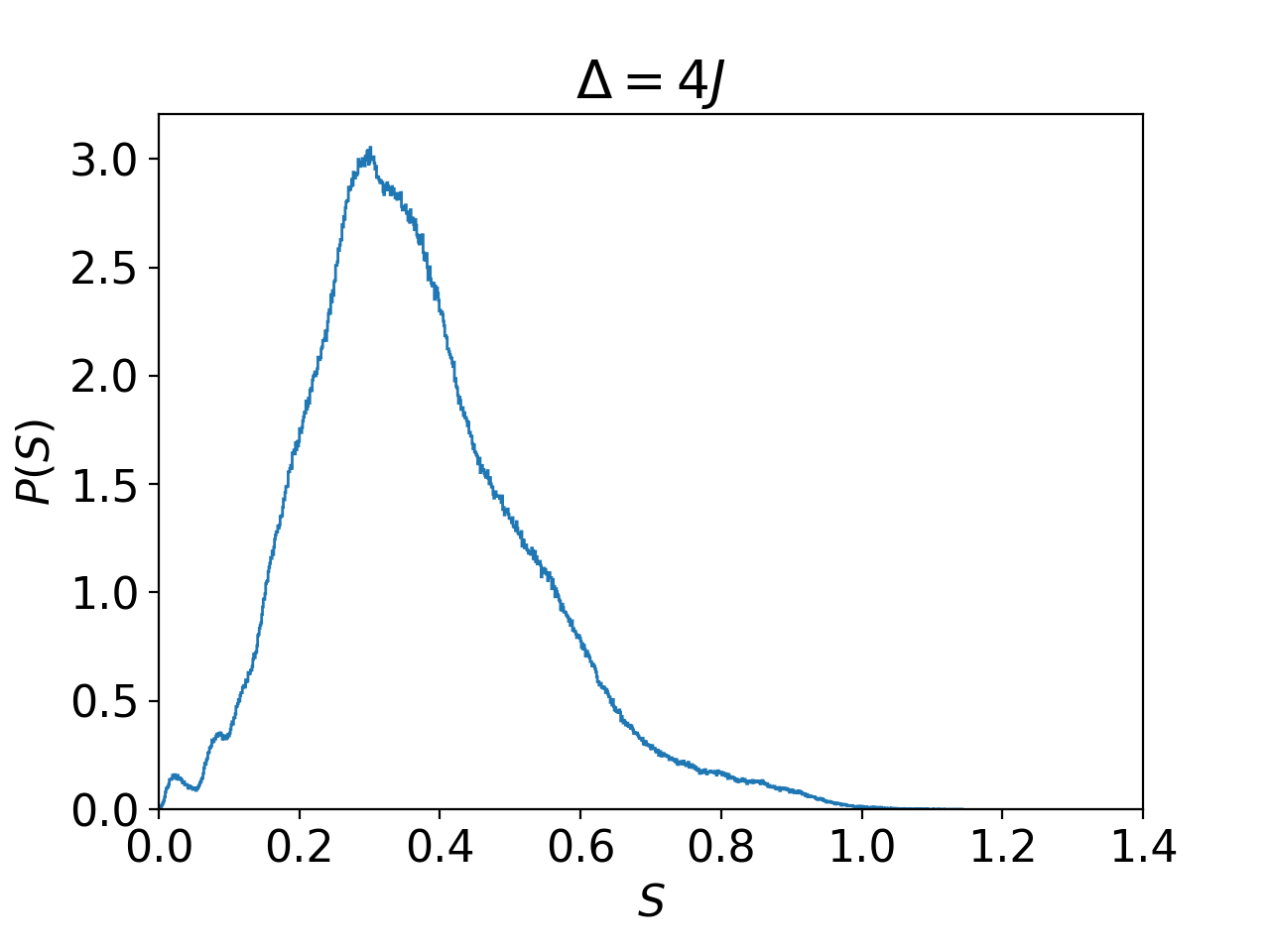} \
\includegraphics[width=0.32\textwidth  ]{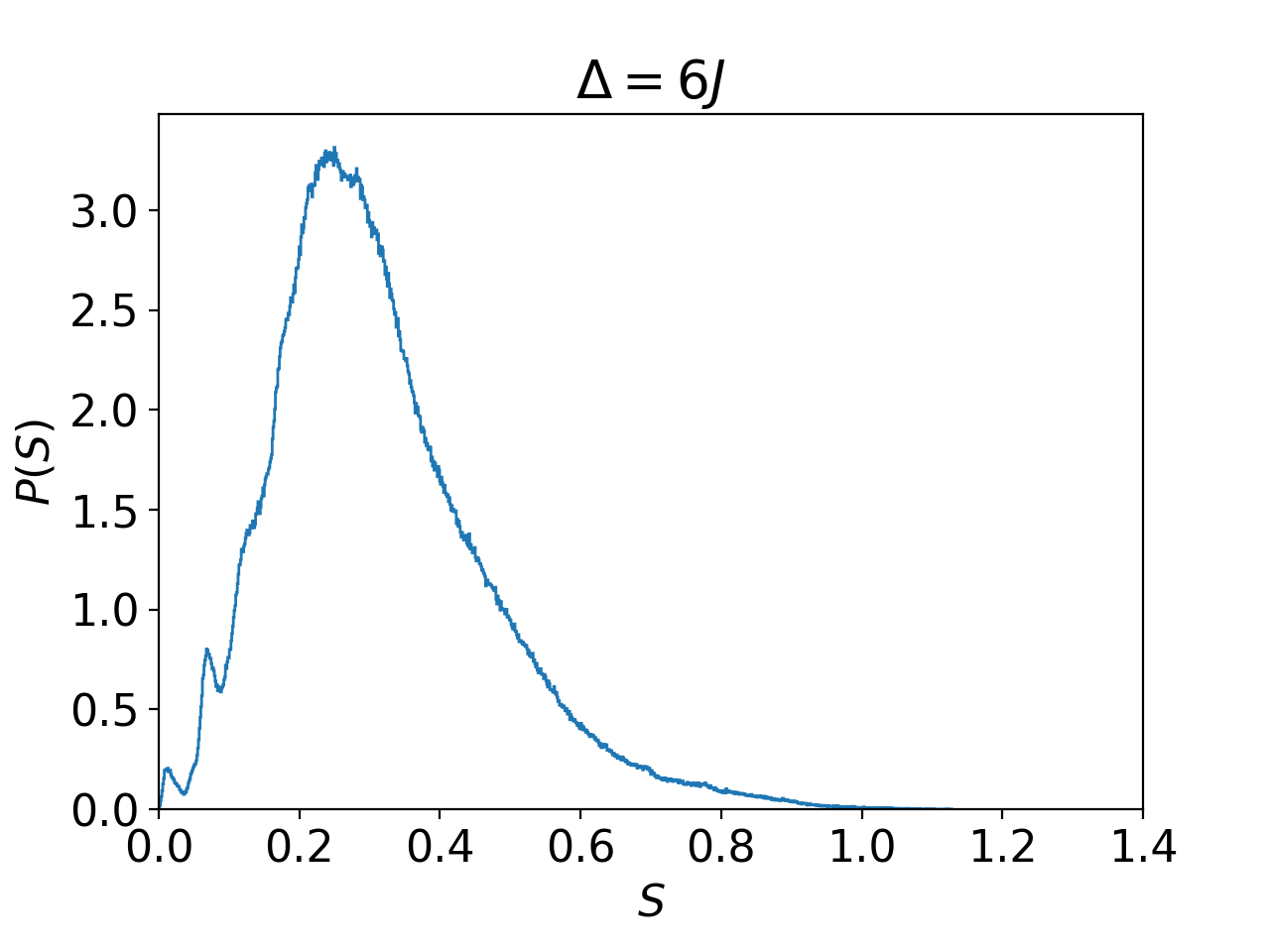} \
\includegraphics[width=0.32\textwidth  ]{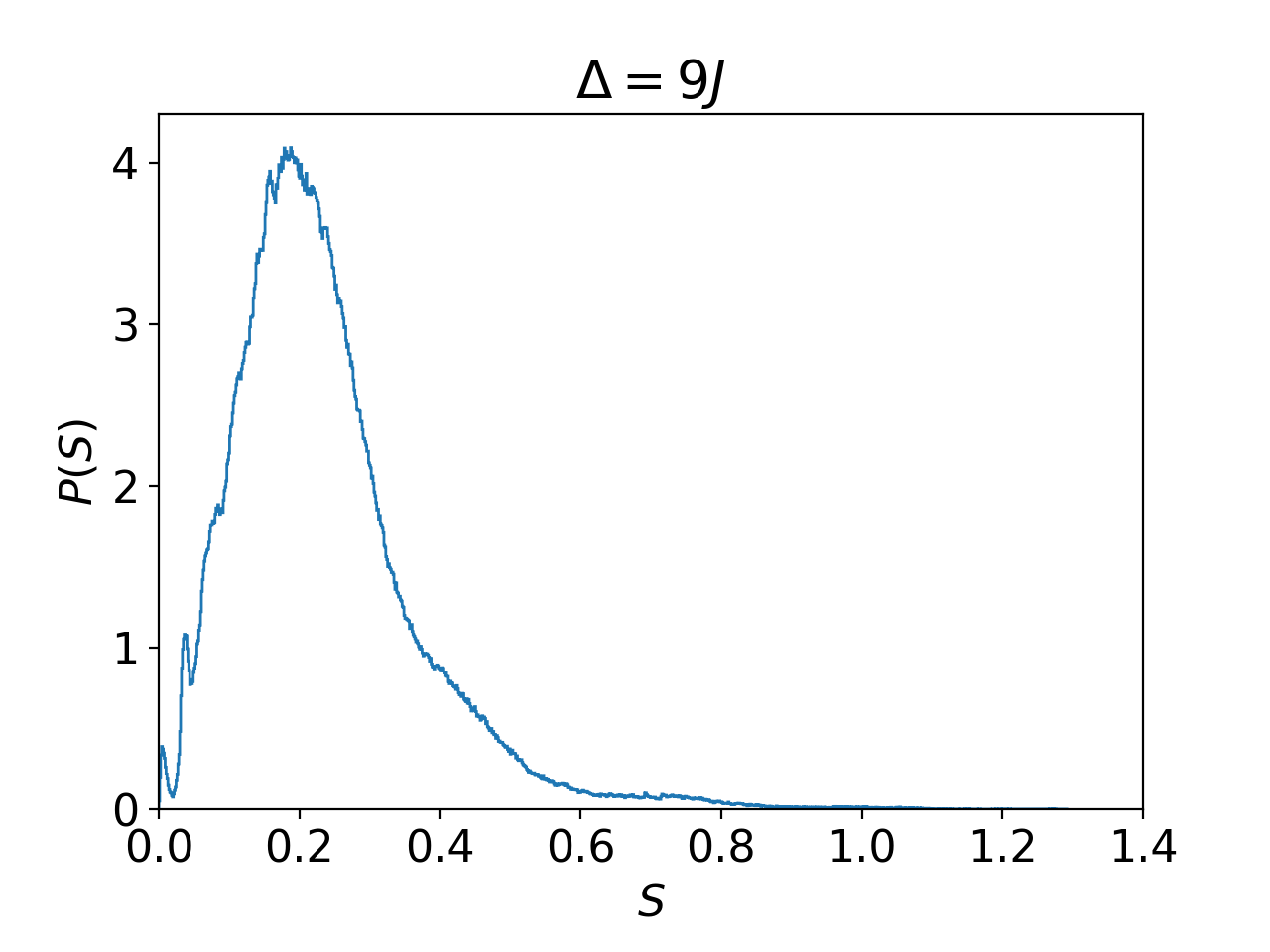} \\
\includegraphics[width=0.32\textwidth  ]{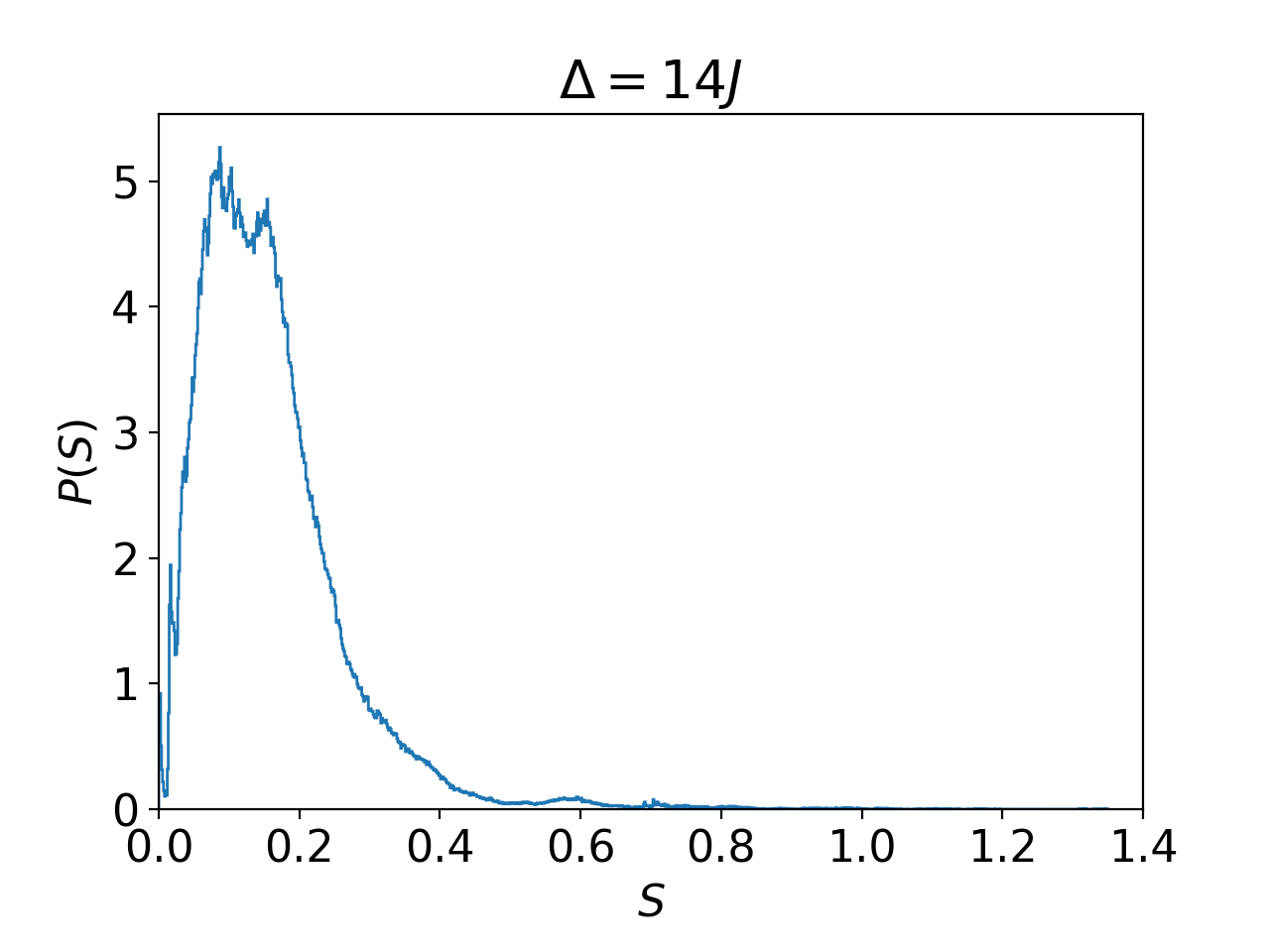} \
\includegraphics[width=0.32\textwidth  ]{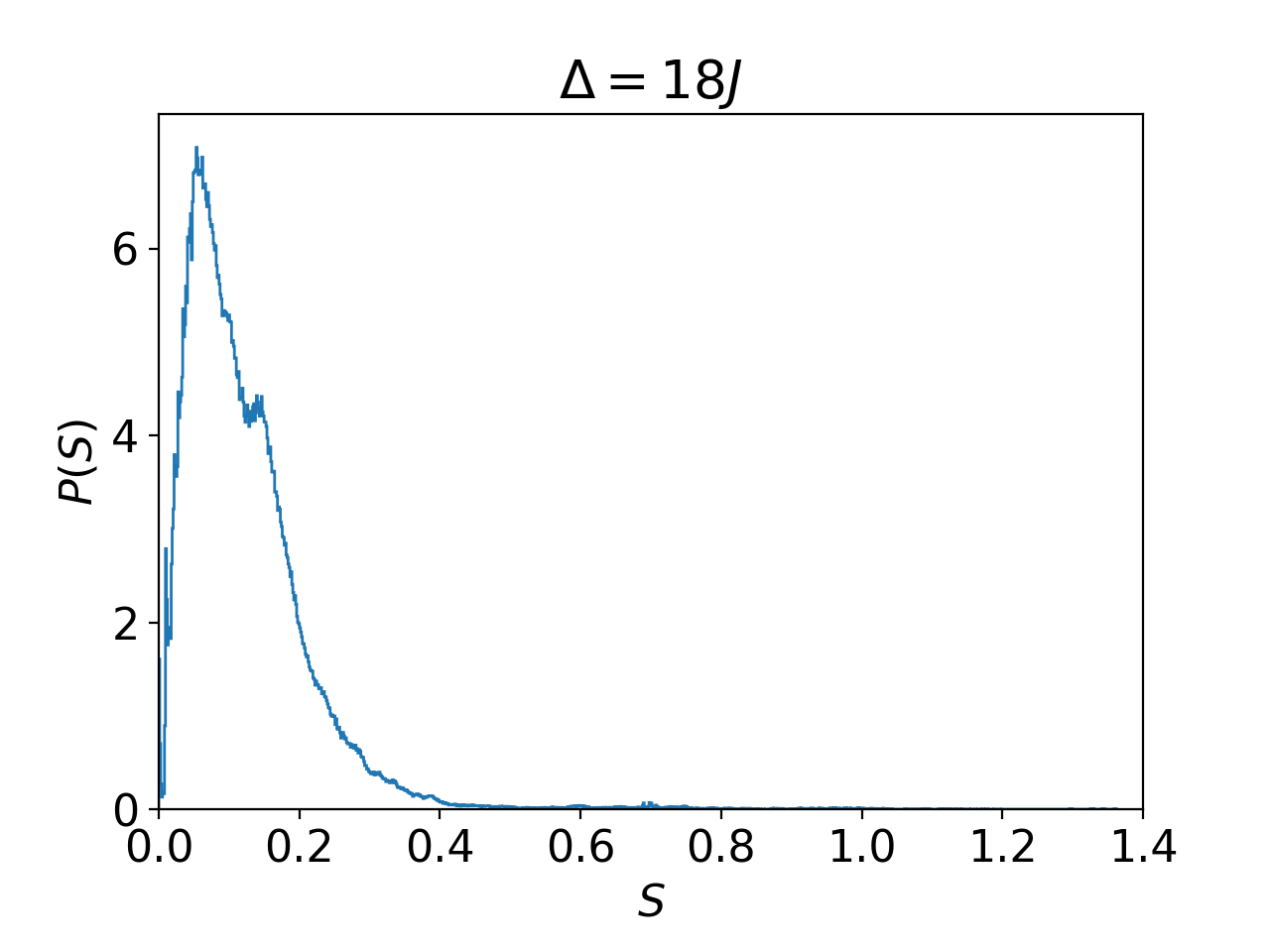}

    \caption{On-site entanglement entropy probability densities from left to right for $\Delta/J = 4,6,9$ (top) and $14,18$ (bottom) for the optimized quantum circuits from the main text. The probability densities correspond to the concatenation of distributions for different lattice sites and disorder realizations. We sampled over $5000$ randomly chosen approximate eigenstates for each lattice site and disorder realization.} \label{fig:histrograms}
\end{figure*}

Here, we present our results of the distributions of on-site entanglement entropies as another signature of the transition near $\Delta = 9J$~\cite{2DMBL}. Specifically, we collected the full distributions of on-site entanglement entropies of the optimized quantum circuits from the main text for $\Delta/J = 4,6,9,14,$ and $18$. In Fig.~\ref{fig:histrograms} we show the corresponding probability distributions sampled over $5000$ approximate eigenstates per disorder realization and lattice site, with the results for all lattice sites and disorder realizations combined. 

We find that a multimodal distribution develops as $\Delta$ is increased to $9J$,  and that as $\Delta$ is increased further, the peaks start to merge, very similar to the results of Ref.~\cite{2DMBL}. The presence of a significant fraction of (approximate) eigenstates with small entanglement and with larger entanglement [the maximum is $\ln(4) \approx 1.386$] around $\Delta = 9J$ is characterisitc of the phase transition point, where there are both area and volume law-entangled eigenstates~\cite{Luitz2015}, which our tensor network approach partially captures~\cite{2DMBL}. However, unlike Ref.~\cite{2DMBL}, we also observe a third peak at $S \approx 0.04$, which is probably caused by the additional degree of freedom in our system.

\begin{figure*}[b]
\centering
\includegraphics[width=0.2\textwidth  ]{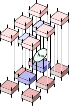}
    \caption{Range of non-trivial action (causal cone) of $\tilde U^\dg h_m \tilde U$ for constituting unitaries of length $\ell = 2$. $\tilde U^\dg h_m \tilde U$ acts as the identity on all other sites.} \label{fig:causal_cone}
\end{figure*}

\section{8. Relation between charge-density-wavelength and the quantum circuit gate range} 

We now provide an argument why our approach reproduces the phase transition point found in charge-density-wave experiments with fermions, while the transition point computed in Ref.~\cite{2DMBL} is significantly higher than the one determined in ``half-moon'' experiments with bosons~\cite{Choi1547}:  
An important aspect of our method is that it describes approximate, strictly short-range LIOMs $\tau_i^z$ (cf. Fig.~\ref{fig:tau}b) rather than the exact integrals of motion of the Hamiltonian $H$. 
Thus, the transition point we find is not that of the original Hamiltonian, but of an auxiliary Hamiltonian $\tilde H$, which is exactly diagonalized by the optimized quantum circuit $\tilde U$. $\tilde H$ is thus given by the approximate LIOMs $\tilde \tau_i^z$, 
\begin{align}
\tilde H = c + \sum_i c_i \tilde \tau_i^z + \sum_{i,j} c_{ij} \tilde \tau_i^z \tilde \tau_j^z + \ldots
\end{align}
(we suppress the $\mu = \ \uparrow, \downarrow$ indices here for simplicity). As our optimized quantum circuit $\tilde U$ also approximately diagonalizes $H$, the coefficients $c_{ijk\cdots}$ can be chosen such that $\tilde H \approx H$. Specifically, this can be achieved by setting 
$c_{ijk\ldots} := \frac{1}{4^{N^2}} \Tr\left(H \tilde \tau_i^z \tilde \tau_j^z \tilde \tau_k^z \ldots\right)$. 
Using the fact that $\tilde \tau_i^z = \tilde U \sigma_i^z \tilde U^\dg$ and $H = \sum_m h_m$ (with two-site terms $h_m$), we obtain 
\begin{align}
c_{ijk\ldots} = \frac{1}{4^{N^2}} \sum_m \Tr(\tilde U^\dg h_m \tilde U \sigma_i^z \sigma_j^z \sigma_k^z \ldots),
\end{align}
which is non-zero only if the sites $i,j,k,\ldots$ all lie in one ``causal cone'' defined by any $\tilde U^\dg h_m \tilde U$, whose maximum length is $3 \ell$ (corresponding to a width of $2\ell$), see Fig.~\ref{fig:causal_cone}. Below, we use this fact to upper bound the distance across which the fiducial Hamiltonian $\tilde H$ can propagate a domain wall.

\begin{figure}
\begin{picture}(250,170)
\put(0,0){\includegraphics[width=0.47\textwidth]{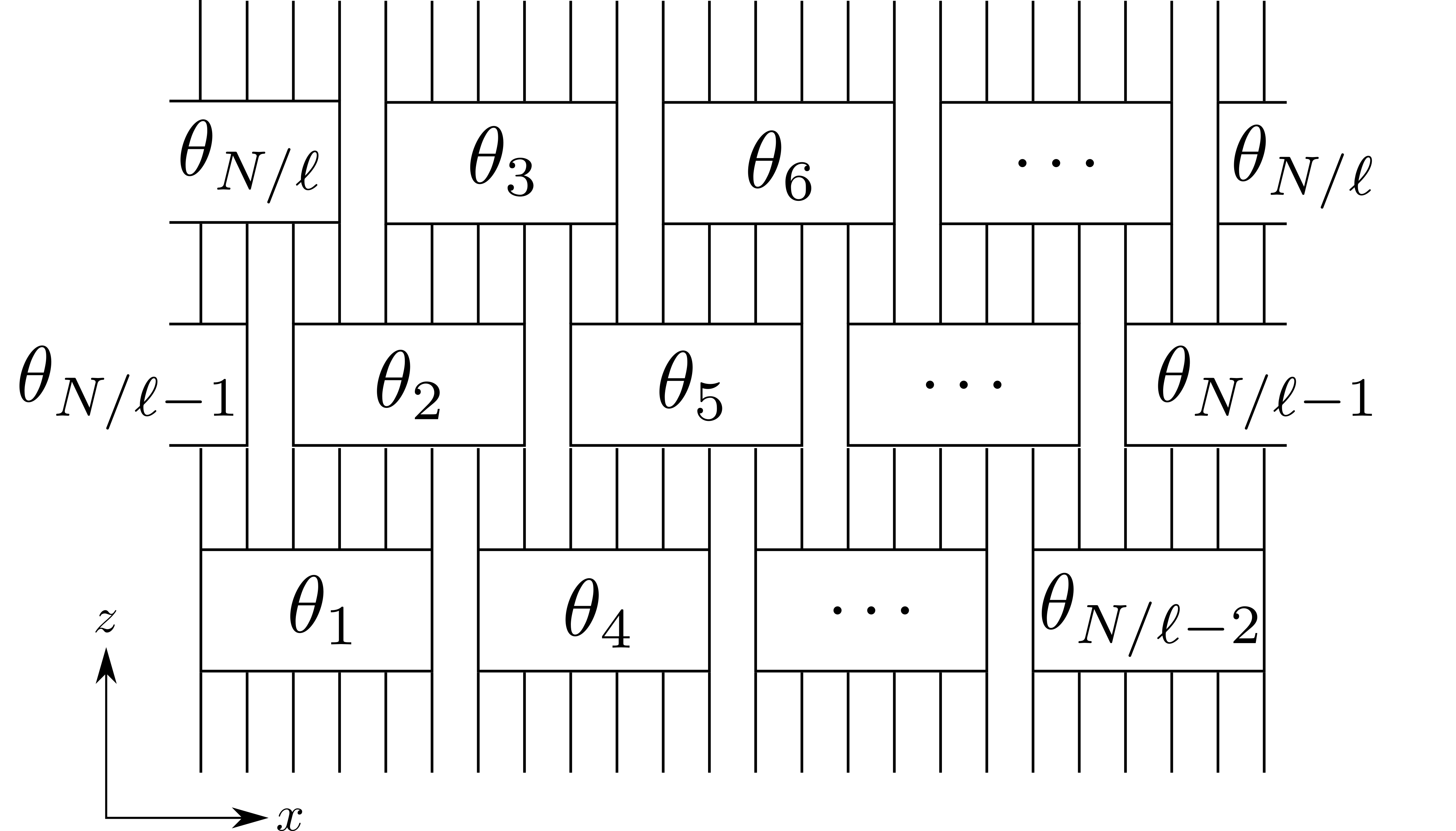}}
\put(-2,170){\textbf{a}}
\end{picture}
\begin{picture}(250,170)
\put(0,0){\includegraphics[width=0.47\textwidth]{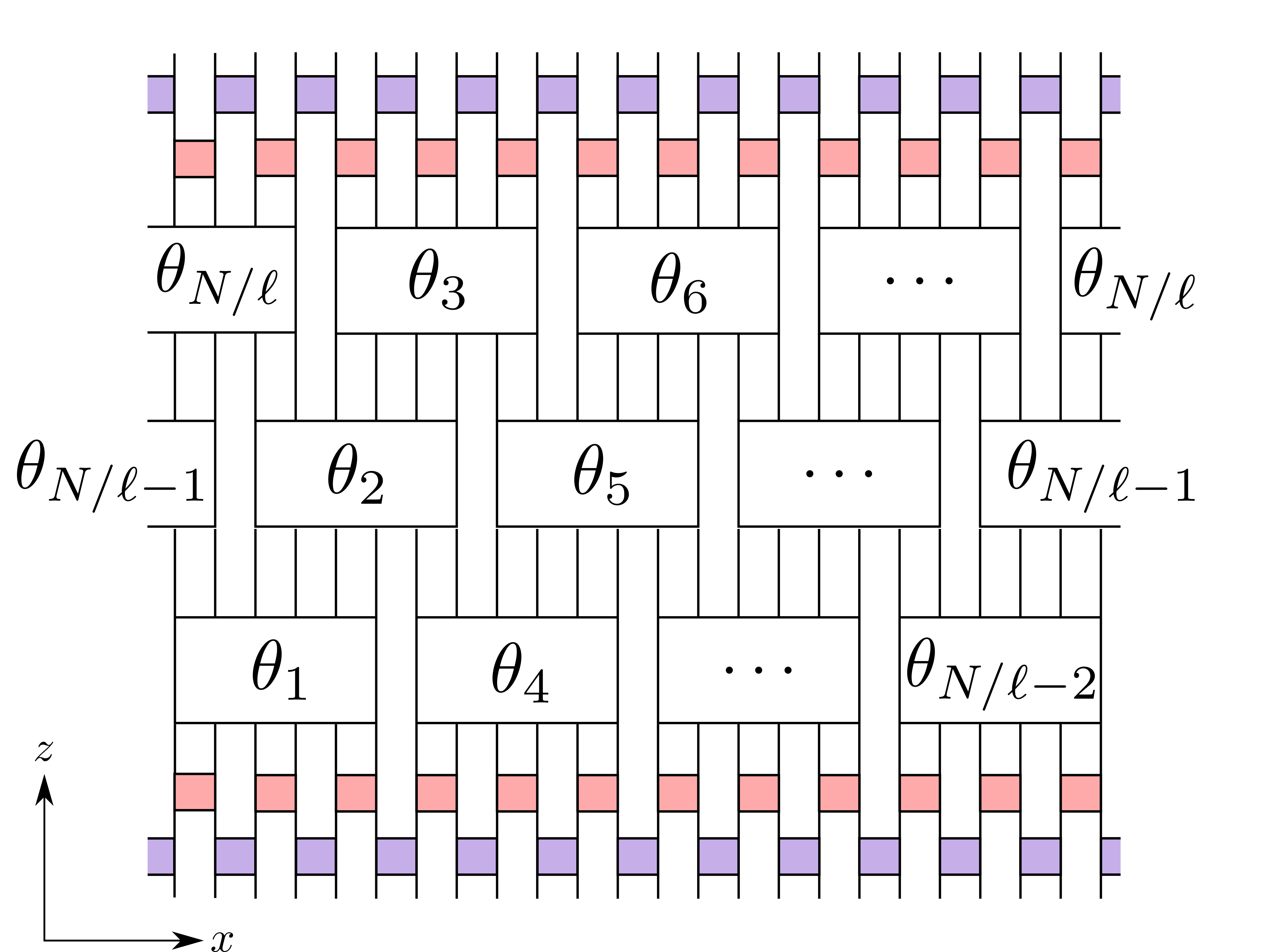}}
\put(-2,170){\textbf{b}}
\end{picture}
    \caption{a: 3-layer quantum circuit representation of $\Theta(t)$ with all unitary gates $\theta_m(t)$ acting on all sites along the $y$-direction for $\ell = 2$. b: Representation of $e^{-i \tilde H t} = \tilde U  \Theta(t) \tilde U^\dg$ after blocking the unitaries of $\tilde U$ (shown in purple and red) along the $y$-direction.} \label{fig:multi-layer}
\end{figure}

\begin{figure}
\centering
\includegraphics[width=0.6\textwidth]{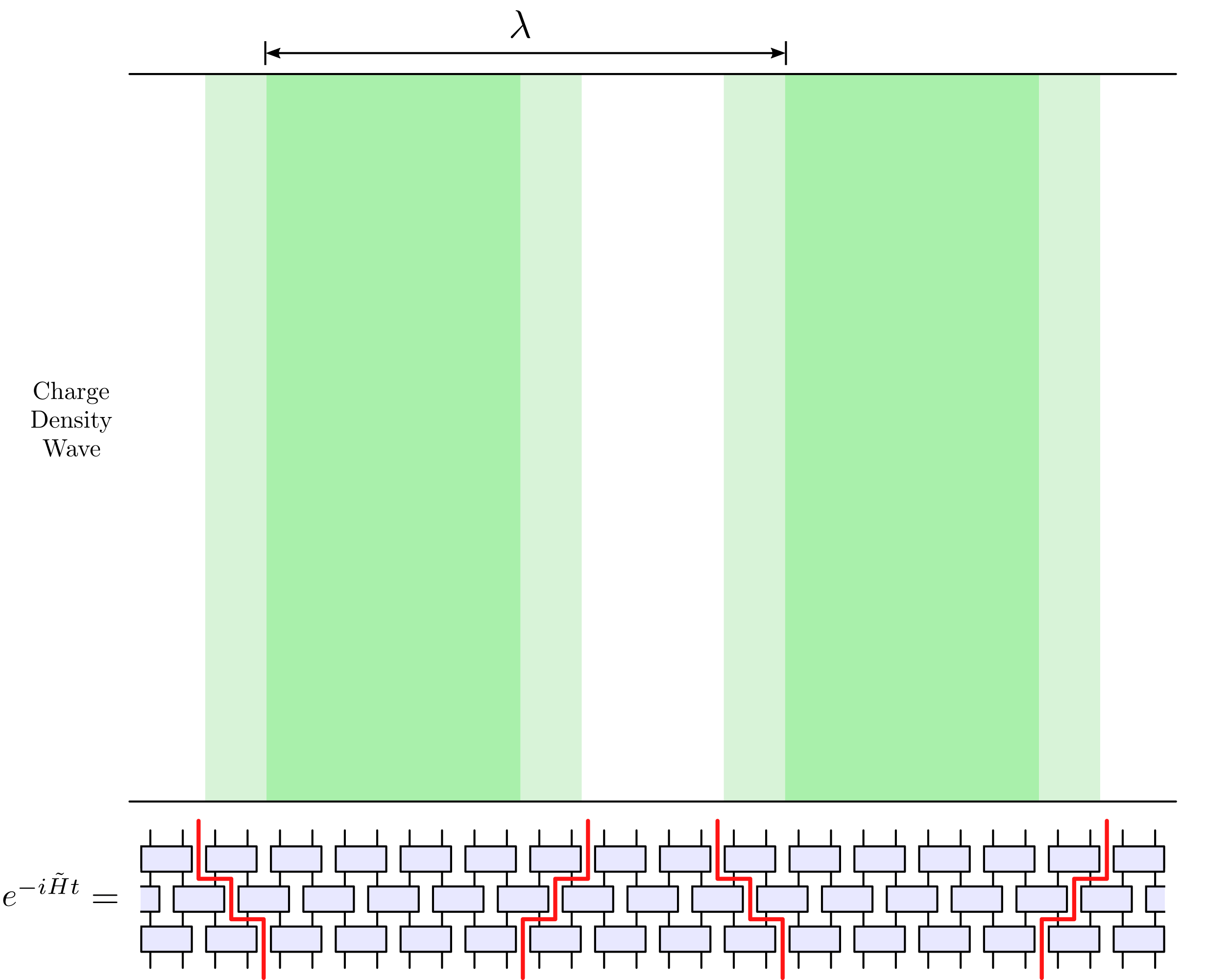}
    \caption{Evolution of a charge-density-wave (top) of wavelength $\lambda$ by $e^{-i\tilde Ht}$ (bottom), which can be written as a one-dimensional 7-layer quantum circuit. 
    For simplicity, we only show 3 layers and unitary gates acting on two sites with causal cones indicated by red lines. 
    The propagated charge-density-wave (top) has regions of high charge density (green) and zero charge density (white).
    In general, the quantum circuit is able to propagate by $\mathcal{O}(\ell)$ columns (shown in light green).}  \label{fig:propagate}
\end{figure}

The time evolution operator corresponding to the auxiliary Hamiltonian takes the form 
\begin{align}
    e^{-i \tilde H t} = \tilde U e^{-i t c} \prod_i e^{-i t c_i \sigma_i^z} \prod_{i,j} e^{-i t c_{ij} \sigma_i^z \sigma_j^z} \ldots \tilde U^\dg, \label{eq:evolution}
\end{align}
where we only keep factors $e^{-i t c_{ijk\ldots} \sigma_i^z \sigma_j^z \sigma_k^z \ldots}$ in the product of Eq. \eqref{eq:evolution} whose coefficients $c_{ijk\ldots}$ are non-vanishing, i.e., where all sites $i,j,k,\ldots$ lie in one causal cone of the type shown in Fig.~\ref{fig:causal_cone}. These factors act non-trivially on plaquettes of at most $3 \ell \times 2 \ell$ sites. The product of these non-trivial factors
\begin{align}
    \Theta(t) := e^{-i t c} \prod_i e^{-i t c_i \sigma_i^z} \prod_{i,j} e^{-i t c_{ij} \sigma_i^z \sigma_j^z} \ldots
\end{align}    
is therefore a fixed-depth quantum circuit representing a diagonal unitary matrix. We now block all of its phase factors acting on columns of $3\ell \times N$ sites together into diagonal unitaries $\theta_m(t)$, $m = 1, \ldots N/{\ell}$, acting non-trivially on all sites with $x$-coordinates $\ell (m-1)+1,\ell (m-1)+2, \ldots, \ell (m+2)$. In this process, $e^{-i t c_{ijk\ldots} \sigma_i^z \sigma_j^z \sigma_k^z \ldots}$ can always be absorbed into at least one $\theta_m(t)$. $\Theta(t)$ can therefore be written as a one-dimensional 3-layer quantum circuit as shown in Fig.~\ref{fig:multi-layer}a (where each unitary gate acts on all sites along the $y$-direction). Hence, $e^{-i \tilde H t} = \tilde U  \Theta(t) \tilde U^\dg$ is a 7-layer one-dimensional quantum circuit, as indicated in Fig.~\ref{fig:multi-layer}b. The latter can propagate particles at most across a distance of up to $d = (2+8+1)\frac{\ell}{2} = 5.5\ell$ sites, cf. causal cone in Fig.~\ref{fig:multi-layer}b (where $\ell =2$). The propagation of an initial charge-density-wave state by $e^{-i \tilde H t}$ is schematically shown in Fig.~\ref{fig:propagate}. As the unitaries of the numerically optimized quantum circuits are in general close to identities and the coefficients $\tilde c_{ijk\ldots}$ exponentially suppressed with the maximum distance between the sites $i,j,k,\ldots$, the Hamiltonian $\tilde H$ is expected to allow for equilibration over much less than $d = 5.5\ell$ columns. Hence, $\ell = 2$ simulations come much closer to the transition point obtained in charge-density-wave experiments~\cite{bordia2017quasiperiodic2D} than with half-moon initializations~\cite{Choi1547}, which is why Ref.~\cite{2DMBL} overestimates the transition point: The ``true'' MBL-to-thermal transition point is upper bounded by the lowest disorder strength for which we find a non-thermalizing initialization, which is expected to be the half-moon configuration~\cite{Choi1547,2D_quantum_bath}, as it requires longer-distance rearrangements of the atoms. 
However, the approach of Ref.~\cite{2DMBL} cannot represent thermalization of such a configuration. Instead, it predicts a higher transition point corresponding to charge-density-waves, which require a larger disorder strength to prevent them from thermalizing.

\end{document}